# Near-direct bandgap WSe$_2$/ReS$_2$ type-II pn heterojunction for enhanced ultrafast photodetection and high-performance photovoltaics

Abin Varghese,[†,‡,¶] Dipankar Saha,[†] Kartikey Thakar,[†] Vishwas Jindal,[§] Sayantan Ghosh,[†] Nikhil V Medhekar,[‡] Sandip Ghosh,[§] and Saurabh Lodha[*,†]

[†]*Department of Electrical Engineering, Indian Institute of Technology Bombay, Mumbai 400076, India*

[‡]*Department of Materials Science and Engineering, Monash University, Clayton, Victoria 3800, Australia*

[¶]*IITB-Monash Research Academy, IIT Bombay, Mumbai 400076, India*

[§]*Department of Condensed Matter Physics and Materials Science, Tata Institute of Fundamental Research, Mumbai 400005, India*

E-mail: slodha@ee.iitb.ac.in

## Abstract

PN heterojunctions comprising layered van der Waals (vdW) semiconductors have been used to demonstrate current rectifiers, photodetectors, and photovoltaic devices. However, a direct or near-direct bandgap at the heterointerface that can significantly enhance optical generation, for high light absorbing few/multi-layer vdW materials, has not yet been shown. In this work, for the first time, few-layer group-6 transition metal dichalcogenide (TMD) WSe$_2$ is shown to form a sizeable (0.7 eV) near-direct bandgap with type-II band alignment at its interface with the group-7 TMD ReS$_2$ through density functional theory calculations. Further, the type-II alignment and photogeneration across the interlayer bandgap have been experimentally confirmed through micro-photoluminescence and IR photodetection measurements, respectively. High optical absorption in few-layer flakes, large conduction and valence band offsets for efficient electron-hole separation and stacking of light facing, direct bandgap ReS$_2$ on top of gate tunable WSe$_2$ are shown to result in excellent and tunable photodetection as well as photovoltaic performance through flake thickness dependent optoelectronic measurements. Few-layer flakes demonstrate ultrafast response time (5 $\mu$s) at high responsivity (3 A/W) and large photocurrent generation and responsivity enhancement at the heterostructure overlap region (10-100×) for 532 nm laser illumination. Large open circuit voltage of 0.64 V and short circuit current of 2.6 $\mu$A enables high output electrical power. Finally, long term air-stability and a facile single contact metal fabrication process makes the multi-functional few-layer WSe$_2$/ReS$_2$ heterostructure diode technologically promising for next-generation optoelectronic applications.

## Introduction

Layered van der Waals (vdW) materials such as graphene, transition metal dichalcogenides (TMDs), and hBN are being studied for applications in photodetection,[1] terahertz spectroscopy,[2] photovoltaics,[3] as well as flexible electronics.[4] The ability to form epitaxy-free heterostructures by vertical stacking of these layered materials has enabled diverse material combinations and device applications.[5-8] Specifically, heterostructures employing vdW semiconductors, with bandgaps in the visible and near-infrared wavelength range, have been used to demonstrate atomically thin pn junctions,[1,6,9-14] light emitting diodes,[9] photovoltaic cells,[6,8,11,13] and memory devices.[15] Although pn junctions using layered semiconductors have also been realized using electrostatic or chemical p and n doping in a single material (homojunctions),[3,16] the approach of conjugating separate p- and n-type materials (heterojunctions) yields atomically abrupt junctions and exhibits efficient electron-hole separation owing to type-II band alignment as seen in WSe$_2$/MoS$_2$,[6,9] BP/MoS$_2$,[1] MoTe$_2$/MoS$_2$,[10,11] ReSe$_2$/ReS$_2$,[12,13] GaTe/MoS$_2$[14] heterostructures. However, a rational selection of the constituent p- and n-type semiconductors should be based on their intrinsic thickness dependent heterointerface electronic bandstructure, band alignment and optical properties, as well as extrinsic stability in the ambient and ability to form low resistance metal contacts. This can lead to the realization of technologically relevant pn junctions with excellent multi-functional (opto)electronic performance as demonstrated through WSe$_2$/ReS$_2$ gated pn heterostructures in this work.

**Selection criteria for p and n materials**

The semiconducting group-6 TMD WSe$_2$, generally found in trigonal prismatic phase,[17] is an indirect bandgap material in its bulk form.[18,19] However, group-7 TMD e.g. 1T phase of ReS$_2$ is distorted octahedral in structure[17,20] and exhibits direct (or, near-direct) bandgap at (or, close to) the Γ point of the Brillouin zone (BZ). Interestingly, unlike the group-6 TMDs, ReS$_2$ exhibits a unique property owing to its distorted structure and weak interlayer coupling- the direct or near-direct nature of its bandgap remains unchanged from monolayer to bulk.[20–22] Closer examination of the bandstructure of group-7 TMDs reveals that the conduction band minimum of ReS$_2$ remains at the Γ point, irrespective of the number of layers.[20,23] But for group-6 TMDs (such as WSe$_2$) the valence band maximum relocates from K to Γ point of the BZ with increasing number of layers.[18,24] It is important to note that the valence band maximum at the Γ point differs in energy only slightly from that at the K point.[18] This gives rise to an increased probability of direct as well as indirect transitions from the Γ and the K valence maxima of WSe$_2$ to the Γ conduction minimum of ReS$_2$ respectively, for a predicted type-II band alignment. The possibility of a direct bandgap transition is not observed in a heterostructure of monolayer WSe$_2$ and monolayer/few-layer ReS$_2$. Hence, interfacing few-layer (FL) flakes of group-6 WSe$_2$ and group-7 ReS$_2$ can significantly enhance internal quantum efficiency through increased heterointerface optical generation, a critical requirement for optoelectronic applications.

Higher photon absorption at the interfacial region of the heterostructure (overlap region) increases photogenerated carrier concentration leading to a larger external quantum efficiency that is essential for photovoltaic applications. Layered vdW materials have large absorption coefficients in the visible range for mono and few layers (nearly $10^5$ cm$^{-1}$ for 600 nm wavelength).[25] While the absorption is ∼ 10% for monolayers,[26] it is significantly higher at ∼ 40% for thicker flakes (∼ 25 nm).[25] This makes FL ReS$_2$ a more appropriate choice for the top, light facing material in the WSe$_2$/ReS$_2$ pn heterostructure stack. Further, ultrathin FL TMDs of group-6 (here, WSe$_2$) show excellent modulation of the Fermi level with applied gate voltage,[27] in contrast to FL ReS$_2$.[28] Controlled variation in the quasi-Fermi level difference between the p and n materials with gate bias can yield tunable open circuit voltages ($V_{OC}$).[29] Hence, FL WSe$_2$ is a more appropriate choice for the bottom, gate facing material in the WSe$_2$/ReS$_2$ pn heterostructure stack. Besides gate-tunable $V_{OC}$, in line with organic solar cells and also shown through atomistic simulations in this work, the large difference between the conduction band minimum of ReS$_2$ and the valence band maximum of WSe$_2$ is expected to result in a large $V_{OC}$ value.[10,30]

Previous studies have shown that Pt contacts on ReS$_2$[31] and WSe$_2$[27] demonstrate n- an p-type transport, respectively. A single metal (Pt) for contacting both WSe$_2$ and ReS$_2$ requires a single lithography-metal deposition-liftoff process sequence. Several reports of pn heterojunctions with dissimilar metal electrodes for the p- and n-type layers (cf. Table 1) require additional fabrication steps. Further, WSe$_2$ and ReS$_2$ are both known to show long term stability in air without any capping layer.[12,27] It is important to note that capping layers have been shown to enhance ambient stability in highly unstable materials such as black phosphorus. However, unlike electronic transport, capping layers can limit optoelectronic performance due to optical absorption losses. In summary, a combination of (i) high optical generation due to the near-direct bandgap at FL group-6/group-7 TMD interface, (ii) high optical absorption in thick TMD layers, (iii) direct bandgap of FL ReS$_2$ and high gate tunability of FL WSe$_2$, (iv) single metal for p and n contacts, and, (v) air stability of uncapped WSe$_2$ and ReS$_2$, can enable excellent optoelectronic performance and ease of fabrication in the type-II pn heterostructure, of a FL flake of ReS$_2$ (group-7) stacked on top of a gated FL flake of WSe$_2$ (group-6), as reported in this work.

Table 1 compares electrical, photodetection and photovoltaic parameters of published vdW pn heterostructures with the FL/FL WSe$_2$/ReS$_2$ heterostructure diode reported in this work, highlighting its versatile and superior optoelectronic performance. It demonstrates excellent pn diode action with a gate tunable current rectification ratio (*RR*) reaching $10^5$. High responsivity (*R* = 3 A/W), large photo-amplification ($10^6$), a 10-to-100 times enhancement of photocurrent, and hence responsivity, at the heterostructure overlap region are consistent with the near-direct type-II band alignment at the heterointerface, calculated using density functional theory and confirmed through temperature and flake thickness dependent micro-photoluminescence (PL) measurements. Moreover, the interlayer bandgap allows infrared (IR) photodetection in energy range much lower than the bandgaps of the constituent WSe$_2$ and ReS$_2$. The ultrafast photoswitching (rise ($\tau_r$) and fall ($\tau_f$) times of ~5 $\mu$s) shows efficient charge separation across the heterojunction and reinforces the predicted large type-II band offsets. Further, the large $V_{OC}$ of 0.64 V is in close agreement with the calculated interlayer bandgap value of 0.7 V. Flake thickness dependent optoelectronic measurements justify the choice of FL flakes to achieve improved photodetection and photovoltaic metrics. Finally, long term photoswitching stability over months in ambient storage conditions, and a single contact metal process highlight the technological relevance of this work.

Table 1: Key electrical, photodetection and photovoltaic parameters of published vdW pn heterojunctions and the FL/FL WSe$_2$/ReS$_2$ heterostructure diode reported in this work.

| Ref. | p/n | M(p)-M(n) | G(p)-G(n) | Stable | Thickness (nm) | Electrical RR | $\eta$ | Photodetection R (A/W) | $\tau_r$ (ms) | Photovoltaic $V_{OC}$ (V) | $I_{SC}$ (nA) |
|---|---|---|---|---|---|---|---|---|---|---|---|
| 6 | WSe$_2$/MoS$_2$* | Al-Pd | 6-6 | Yes | 0.8/0.8 | - | - | 0.01 | - | 0.5 | 70 |
| 9 | WSe$_2$/MoS$_2$* | Au | 6-6 | Yes | 1.6/0.8 | - | - | - | < 0.1 | 0.27 | 220 |
| 1 | BP*/MoS$_2$ | Au | -6 | No | 11/0.8 | 10$^5$ | 2.7 | 0.4 | - | ∼ 0.3 | ∼ 15 |
| 12 | ReSe$_2$/ReS$_2$* | Pd | 7-7 | Yes | 50.4/3.5 | - | - | 10$^5$ | 4 | - | - |
| 10 | MoTe$_2$*/MoS$_2$ | Pt-Au | 6-6 | Yes | 3/3 | 10$^3$ | 1.3 | 0.3 | 25 | 0.3 | 200 |
| 29 | GeSe/WSe$_2$* | Au | -6 | Yes | 52/6.5 | 10$^4$ | 1.7 | 10$^3$ | - | 0.7 | 3 |
| 32 | GaSe*/InSe | Gr | - | No | 19/13 | 10$^5$ | - | 350 | 2 × 10$^{-3}$ | - | - |
| 11 | MoTe$_2$*/MoS$_2$ | Pd-Au | 6-6 | Yes | 3.3/7 | 80 | - | 0.3 | 0.06 | 0.51 | 1 × 10$^3$ |
| 13 | ReSe$_2$/ReS$_2$* | Pt-Ti | 7-7 | Yes | 48/64 | 364 | - | 0.02 | 400 | 0.17 | 0.05 |
| This work | WSe$_2$/ReS$_2$* | Pt | 6-7 | Yes | 34/46 | 10$^3$- 10$^5$ | 2 | 3 | 5 × 10$^{-3}$ | 0.64 | 2.6 × 10$^3$ |

G ∼ TMD group, M ∼ Metal, Gr ∼ Graphene, $\eta$ ∼ ideality factor, * denotes the material on top

## Atomistic Model

An atomistic study employing first-principles based density functional theory (DFT) calculations was carried out to gain insights into fundamental electronic properties of the WSe$_2$/ReS$_2$ vdW heterostructures. The DFT calculations were performed using the software package 'QuantumATK'.[33,34] To capture bulk intrinsic properties of the WSe$_2$/ReS$_2$ heterostructure, 4 layers (4L) of WSe$_2$[35] and 2 layers (2L) of distorted 1T-ReS$_2$[36] were considered (as illustrated in Figure 1a). The indirect bandgap of 4L WSe$_2$ was found to be 1.24 eV, whereas the direct bandgap of 2L ReS$_2$ was 1.40 eV (Figure 1b). Details of the calculations can be found in the 'Methods' section as well as in Supporting Information, S1. It should be noted that the bandgaps calculated for 4L WSe$_2$ and 2L ReS$_2$ are consistent with those of bulk WSe$_2$ and ReS$_2$.[18,37,38] To further compare the calculated layer-dependent bandgap values of WSe$_2$ and ReS$_2$ with previously reported theoretical and experimental results, tables T1, T2 and figures S1.2, S1.3 have been included in the Supporting Information.

Calculated bandstructure in Figure 1b (i) depicts the conduction band minimum for 2L ReS$_2$ ($CB_{min\_R}$) at the Γ point. Comparing this with monolayer DFT results (Supporting Information, Figure S1.1), it can be noticed that the $CB_{min\_R}$ is near the Γ point irrespective of the number of layers.[23,39] However, for WSe$_2$ the valence band maximum ($VB_{max\_W}$) relocates from the K point to the Γ point of the BZ (Figure 1b (ii)), as the number of layers is increased from a monolayer to four layers. The $VB_{max\_W}$ calculated at the Γ point is only 0.08 eV above than that at the K point. Such a trend in valence band dispersion has already been shown in previous studies.[18,24] Figure 1c shows the computed bandstructure of the composite 4L/2L WSe$_2$/ReS$_2$ vdW heterostructure, along with the projections of element-wise contribution. Structural transformation of constituent materials into an optimized composite supercell gives rise to type-II band alignment with conduction band minimum at the $\Gamma_S$ point and valence band maximum at the K$_S$ point of the BZ. Moreover, the composite bandstructure is not just a superposition[40] of the bands originating from 2L ReS$_2$ (Figure 1b (i)) and 4L WSe$_2$ (Figure 1b (ii)), since it incorporates several quantum mechanical and structural effects like charge redistribution and percentage of strain on each material. The conduction band minimum due to ReS$_2$ forms a 0.68 eV indirect bandgap ($\Gamma_S$ to K$_S$) with the valence band maximum contributed by WSe$_2$. However, the indirect bandgap is very close to the direct bandgap value (0.7 eV, at $\Gamma_S$). The possible optical transitions in the heterojunction are illustrated by double-headed arrows in Figure 1d, where the direct and indirect transition probabilities are represented by T$_{WR\_D}$ ($\Gamma_S$ to $\Gamma_S$) and T$_{WR\_I}$ (K$_S$ to $\Gamma_S$), respectively. It should be noted that for optical generation across the heterointerface, the direct bandgap transition T$_{WR\_D}$ will not require phonon assisted momentum conservation, unlike T$_{WR\_I}$. Hence, this near-direct[41] electronic structure can lead to significantly large optical transitions at $\Gamma_S$ which may contribute to enhanced photogeneration. This strongly correlates to the high responsivity obtained from the heterojunction in our focused laser measurements.

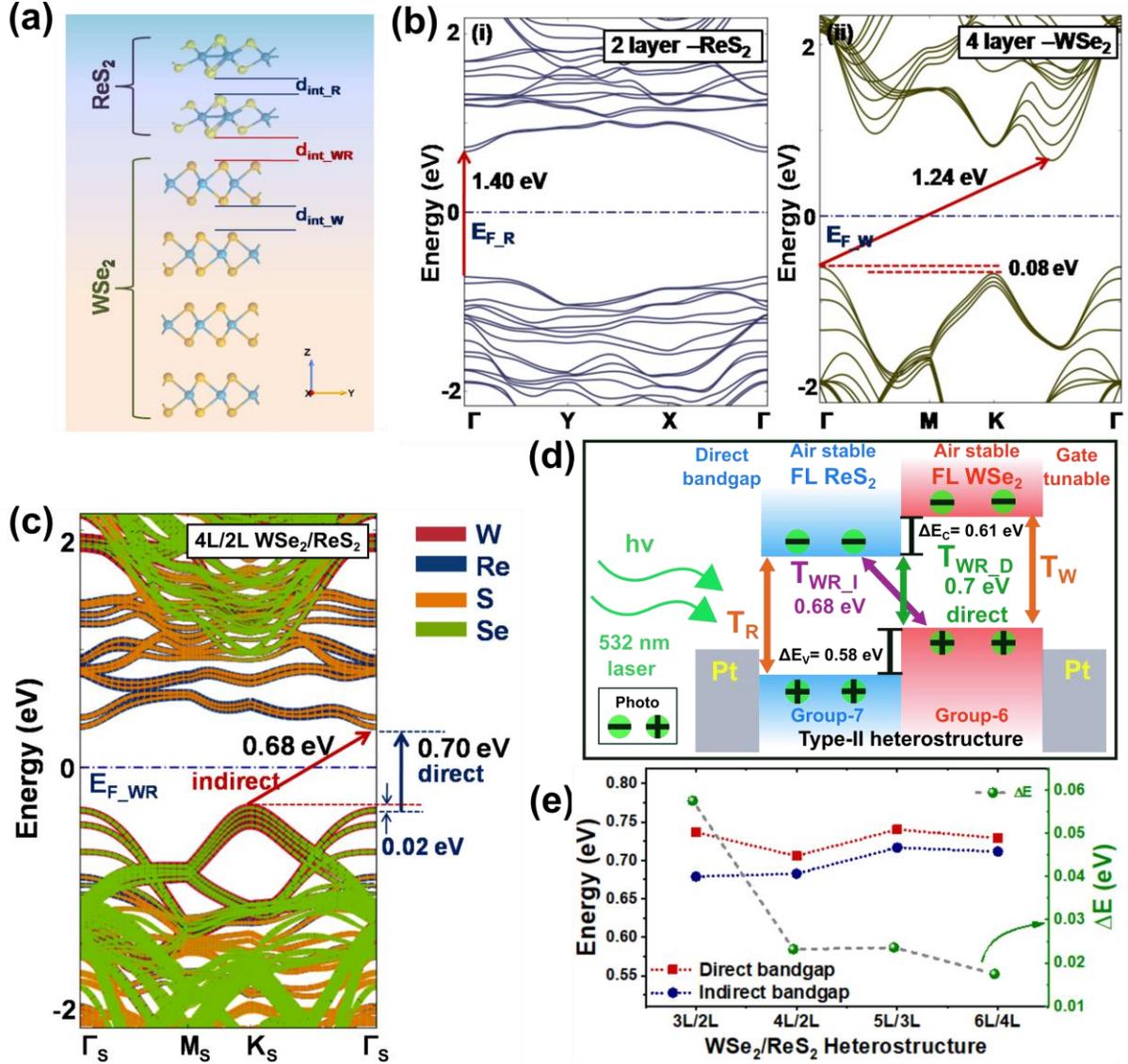

Figure 1: (a) Atomistic representation of 4L/2L WSe$_2$/ReS$_2$ vdW heterostructure, where $d_{int\_R}$ (∼3.2 Å), $d_{int\_W}$ (∼3.3 Å), and $d_{int\_WR}$ (∼3.4 Å) are the optimized equilibrium distances. (b) Bandstructure of 2L ReS$_2$ and 4L WSe$_2$. (c) Projected bandstructure of 4L/2L WSe$_2$/ReS$_2$ vdW heterostructure showing element-wise contributions. $E_{F\_R}$, $E_{F\_W}$, and $E_{F\_WR}$ denote the Fermi level positions of 2L ReS$_2$, 4L WSe$_2$, and WSe$_2$/ReS$_2$ vdW heterostructure, respectively. The direct and indirect bandgap values are only 0.02 eV apart. (d) Schematic illustrating the near-direct type-II band alignment of the WSe$_2$/ReS$_2$ heterostructure. Here, the $\Delta E_C$ and $\Delta E_V$ values for the 4L/2L structure are 0.61 eV and 0.58 eV, respectively. The direct and indirect transition probabilities across the heterojunction are marked as T$_{WR\_D}$ and T$_{WR\_I}$, respectively. T$_W$ is the transition probability in WSe$_2$ and T$_R$ in ReS$_2$. (e) Calculated direct and indirect bandgaps as well as the corresponding energy difference (ΔE) values of the various WSe$_2$/ReS$_2$ heterostructures.

Furthermore, a closer look at Figure 1c shows a type-II bandstructure necessary for efficient separation of the photogenerated electrons and holes[42,43] and can offer strong interlayer coupling via tunnelling of the majority carriers.[6,42] This is further reinforced by strong photoluminescence quenching (Figure 3) and the region-wise photoresponse (Figure 5d) indicates maximum photoresponse at the WSe$_2$/ReS$_2$ interface region. Figure 1d also which illustrates a schematic representation of the projected bandstructure shown in Figure 1c, where the conduction and valence band offsets ($\Delta E_C$ and $\Delta E_V$) are 0.61 eV and 0.58 eV, respectively. The near-direct interface bandgap (0.7 eV) enhances photogeneration and the large values of $\Delta E_C$ and $\Delta E_V$ favour easy and fast dissociation of the photogenerated electrons and holes thereby improving the photoresponsivity.[6,42,43] The calculated sizeable bandgap value of 0.7 eV can also result in a relatively large $V_{OC}$.[30]

To further investigate the band alignment in thicker WSe$_2$/ReS$_2$ heterostructures, atomistic models were designed using 5 layers of WSe$_2$ and 3 layers of ReS$_2$ (5L/3L WSe$_2$/ReS$_2$) and 6 layers of WSe$_2$ and 4 layers of ReS$_2$ (6L/4L WSe$_2$/ReS$_2$). Apart from that, a thinner 3L/2L WSe$_2$/ReS$_2$ heterostructure has also been considered for comparison. The electronic bandstructures for 3L/2L WSe$_2$/ReS$_2$, 5L/3L

WSe$_2$/ReS$_2$, and 6L/4L WSe$_2$/ReS$_2$ heterointerfaces are shown in Figure S1.3 of Supporting Information. The direct and indirect bandgaps as well as their energy difference (ΔE) for the thicker and thinner heterostructures are compared with the aforementioned 4L/2L case in Figure 1e. Although the thick (4L/2L and above) structures exhibit indirect bandgaps around 0.7 eV, their ΔE values always remain close to or less than 0.02 eV, rendering a near-direct nature[41] to the bandstructure. Figure 1e justifies that the 4L/2L case can be used to represent the bandstructure of the thicker FL/FL WSe$_2$/ReS$_2$ heterostructure used for the optoelectronic measurements (described in the next section).

## Results and Discussion

### Device fabrication and physical characterization

FL WSe$_2$ and ReS$_2$ flakes were exfoliated from bulk crystals onto polydimethylsiloxane (PDMS) films. The WSe$_2$ flake was transferred first onto a 280 nm SiO$_2$/Si substrate. Next, the FL ReS$_2$ flake was aligned and transferred over WSe$_2$ ensuring a good overlap for the heterojunction. A schematic of the heterostructure showing the ReS$_2$ flake on top of WSe$_2$ is shown in Figure 2a. False colored scanning electron microscopy (SEM) image of the heterojunction is shown in Figure 2b and the contacts are labelled in the inset. An optical microscope image is available in Figure S2 of Supporting Information. Raman spectra from the individual materials as well as the overlap region are shown in Figure 2c. For FL ReS$_2$, Raman peaks at 161 and 212 cm$^{-1}$ are identified as the E$_{2g}$ (in-plane) and A$_{1g}$ (out-of-plane) vibration modes, respectively.[28] FL WSe$_2$ exhibits A$_{1g}$ and 2LA(M) (second order Raman mode) peaks at 249 and 257 cm$^{-1}$ respectively.[44] Characteristic peaks corresponding to both materials are distinctly seen in the overlap region indicating good interface quality. Further, the thickness of each individual flake obtained by atomic force microscopy (AFM) is marked in the AFM image in Figure 2d. This device will be referred to as FL/FL WSe$_2$/ReS$_2$ heterostructure.

To examine the optical properties of the WSe$_2$/ReS$_2$ heterostructure, micro-PL studies were carried out using 532 nm laser with a spot diameter of ∼4 $\mu$m. In addition to the aforementioned FL/FL structure, two samples with thinner constituents: (1) FL/8L WSe$_2$/ReS$_2$ (65 layers of WSe$_2$ and 8 layers of ReS$_2$) and (2) 2L/2L WSe$_2$/ReS$_2$ (Figure 3a) were studied (AFM data available in Supporting

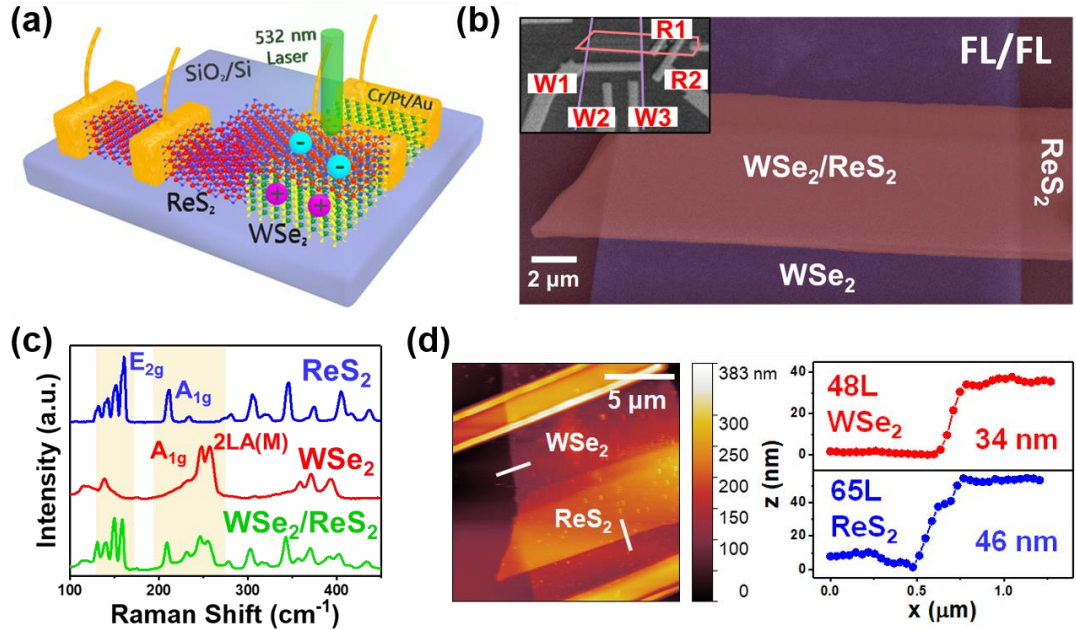

Figure 2: (a) Schematic of the $WSe_2/ReS_2$ heterostructure. (b) Scanning electron microscope image of the fabricated FL/FL heterostructure device. The bottom flake shown in blue is $WSe_2$ and $ReS_2$, in red, is placed on top. The inset shows the contact labels. (c) Raman spectra acquired from individual $WSe_2$ and $ReS_2$ regions as well as from the overlap region showing characteristic peaks corresponding to both materials. (d) Atomic force microscope image of the FL/FL heterojunction area showing flake thicknesses.

Information, S3). The laser was selectively focused on the individual $WSe_2$ and $ReS_2$ regions as well as the $WSe_2/ReS_2$ overlap region. At 300 K, for the relatively thicker FL $WSe_2$, in Figure 3b, the A and I peaks corresponding to the direct (K in conduction band to K in valence band) and indirect (between $\Gamma$ and K in conduction band to $\Gamma$ in valence band) transitions, respectively, and the direct bandgap peak (D) for 8L $ReS_2$ at $\Gamma$ are prominent.[18,20] The characteristic peaks from both materials are significantly quenched in the PL signal from the overlap region. For the thicker FL $WSe_2$, the signal from the indirect bandgap is pronounced, hinting towards the increased possibility of transitions to the $\Gamma$ of the BZ and in turn accumulation of photogenerated holes at $\Gamma$. The direct transitions in $ReS_2$ also occur at $\Gamma$. For the thinner 2L/2L $WSe_2/ReS_2$ heterostructure, at 300 K, the A and I peaks are merged, whereas at 25 K, the A peak shows a blueshift[18] owing to the typical increase in the direct bandgap of a semiconductor with decrease in temperature. The PL from 2L layer $ReS_2$ was very weak at room temperature and hence the measurement at 25 K was performed. In the 2L/2L heterostructure also, it was observed that the PL intensity was significantly reduced at the overlap region owing to efficient transfer of the laser induced photogenerated carriers across the heterojunction[6,45] (electrons to ReS$_2$ and holes to WSe$_2$) indicating a type-II band alignment, in good agreement with the DFT predictions and consistent with many previous reports.[6,45,46] The PL quenching data from a FL/FL heterostructure is available in S4 of Supporting Information. The absence of a PL signal due to recombination of interlayer excitons, if any, in the PL spectra captured by the Si-based detector (detection range upto 1.1 eV) employed in this study reinforces the possibility of a lower energy (sub-1.1 eV) type-II interlayer bandgap.

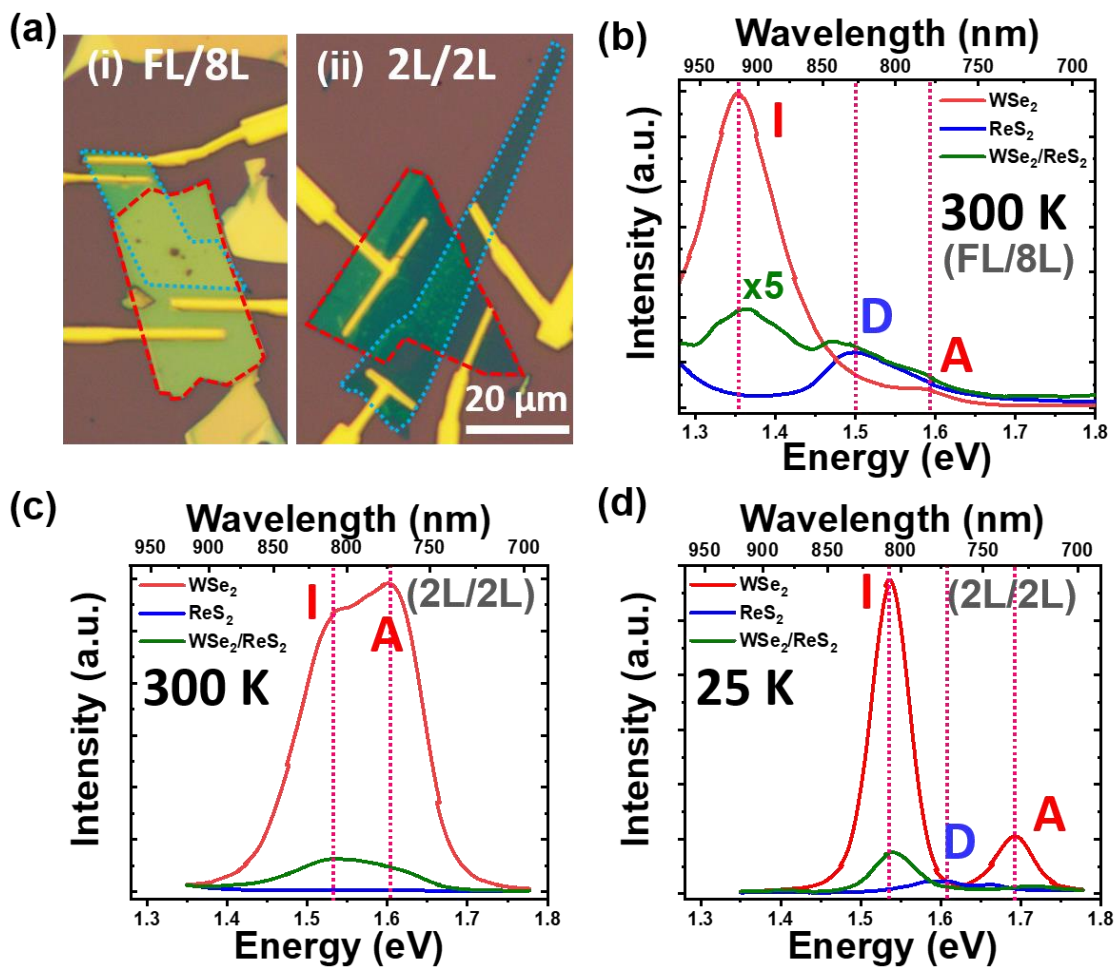

Figure 3: (a) Optical microscope images of (i) FL/8L and (ii) 2L/2L WSe$_2$/ReS$_2$ heterostructures. WSe$_2$ flakes are marked in red and ReS$_2$ in blue. (b) Room temperature PL signal from individual WSe$_2$ and ReS$_2$ regions as well as the WSe$_2$/ReS$_2$ overlap region of the FL/8L heterostructure showing considerable quenching of PL in the overlap region (multiplied by 5 for clarity). (c) Room temperature and (d) low temperature (25 K) PL data for 2L/2L heterostructure which also show PL quenching in the overlap region.

## Electrical characterization without illumination

For electrical characterization, multiple electrodes were fabricated on the heterostructures using electron beam lithography followed by sputtering of a Cr (< 5 nm)/Pt (50 nm)/Au (50 nm) metal stack. Pt has been reported to form ohmic contacts to ReS$_2$[31] as well as WSe$_2$.[27] Individual field effect transistor (FET) performance was first analyzed using contacts only on WSe$_2$ (W2-W3) and only on ReS$_2$ (R1-R2). Transfer ($I_D$ – $V_G$) and output ($I_D$ – $V_D$) characteristics of these individual FETs are shown in Figure S5 of Supporting Information. For both FETs, the threshold voltage ($V_{Th}$) was obtained at the maximum transconductance ($g_m$) point in the transfer characteristics and the carrier mobility ($\mu$) was calculated at $V_{Th}$ using the equation: $\mu = \frac{L/W}{C_G V_D}\frac{dI_D}{dV_G}$. Here, $L$ and $W$ are the length and width of the channel, $V_G$ and $V_D$ are the gate and drain voltages, $I_D$ is the drain current, and the oxide capacitance $C_G$ = 1.23× 10$^{-8}$ Fcm$^{-2}$. The WSe$_2$ FET shows excellent modulation of $I_D$ with $V_G$, demonstrating an on/off current ratio of 10$^5$, with $V_{Th}$ = −28 V and $\mu$ = 32 cm$^2$ V$^{-1}$ s$^{-1}$. On the other hand, the ReS$_2$ FET shows a poor gate tunability of $I_D$ and a lower $\mu$ = 3 cm$^2$ V$^{-1}$ s$^{-1}$. FL ReS$_2$ flakes have been shown to exhibit very weak interlayer coupling[20] which could lead to reduced gate dependent charge carrier modulation in thicker flakes. Nevertheless, on/off current ratio of 10$^8$ is shown for a thinner (trilayer) ReS$_2$ FET in Figure S6 of Supporting Information. From the transfer characteristics of the individual FETs, n- and p-type characteristics are clearly observed for ReS$_2$ and WSe$_2$, respectively. Similar transport characteristics were observed for heterostructure devices without the Cr adhesion layer (S7 of Supporting Information). Further, the choice of WSe$_2$ as the bottom, gate facing, electrically tunable layer and FL direct bandgap ReS$_2$ as the top, light-facing, high optical absorption layer is justified.

The diode characteristics were measured across the contacts W1 on WSe$_2$ and R1 on ReS$_2$. $V_G$ dependent output characteristics are shown in Figure 3a. WSe$_2$ was connected to the drain and ReS$_2$ to the source terminal. A maximum *RR*, defined as the ratio of the forward current ($I_F$) to the reverse current ($I_R$) at $V_D$ = ±1 V, of 6×10$^3$ was obtained at $V_G$ = −20 V. On comparing with the Shockley diode equation, $I_D = I_0 \left(\exp\left(\frac{qV_D}{\eta k_B T}\right) - 1\right)$, where $I_0$ is the reverse leakage current, $q$ is electronic charge, $k_B$ is Boltzmann's constant, $T$ is temperature, and $\eta$ is ideality factor, the $\eta$ is approximately 2 for various $V_G$, indicating recombination limited transport across the heterojunction. Figure S8 in the Supporting Information describes a similar device with $RR$ reaching 10$^5$. A significantly large series resistance ($R_s$) of the order of MΩ is extracted from the output characteristics at high $V_D$, which further increases at positive $V_G$ due to reduced hole density in WSe$_2$. The output characteristics of the diode at positive $V_G$ are shown in Figure S9 of Supporting Information.

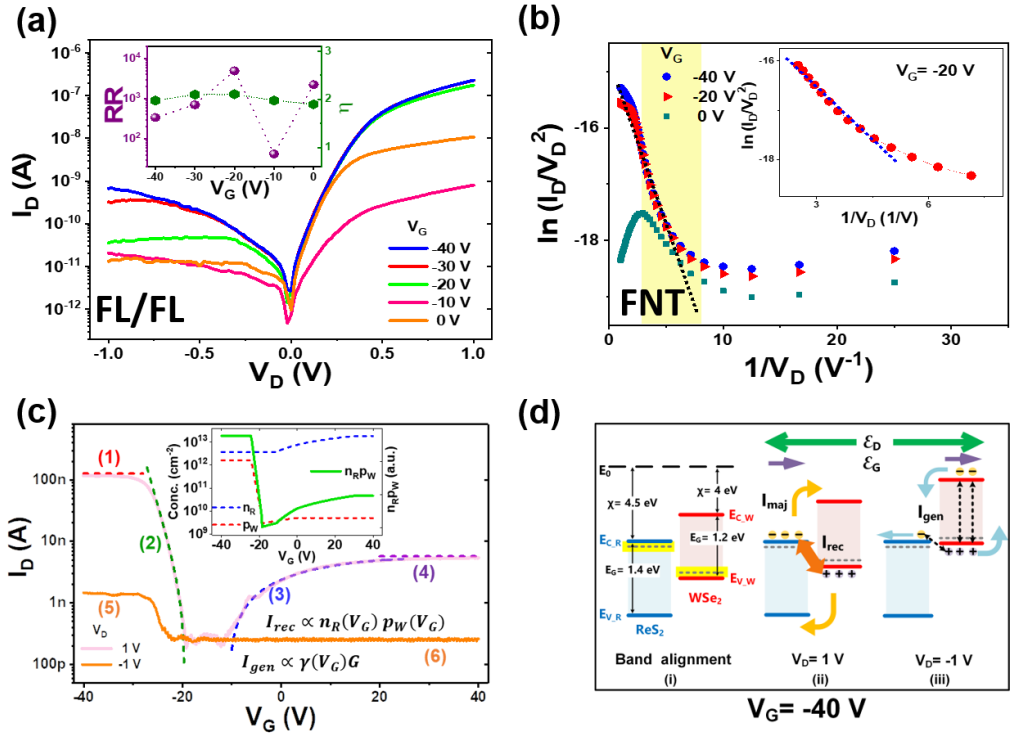

Figure 4: (a) Gate tunable output characteristics of the FL/FL WSe$_2$/ReS$_2$ pn heterojunction across the contacts W1 and R1. Inset shows the extracted values of rectification ratio and diode ideality factor as a function of applied back gate voltage. (b) Plot of $\ln(I_D/V_D^2)$ vs $1/V_D$ showing a linear region with a negative slope (inset) indicating Fowler-Nordheim tunneling transport through the heterostructure. (c) Transfer characteristics of the pn heterojunction for forward and reverse bias conditions with the various regions marked. Inset shows gate voltage dependent extracted hole sheet density in WSe$_2$ ($p_W$) and electron sheet density in ReS$_2$ ($n_R$) along with the product $n_R p_W$ that determines the interlayer recombination current. (d) Energy band diagrams for $V_G = -40$ V: (i) band alignment in WSe$_2$ and ReS$_2$ with their respective Fermi levels defined by the gate voltage, (ii) under forward bias, the dominant charge transport is governed by interlayer recombination of the majority carriers giving rise to $I_{rec}$, and (iii) for reverse bias the reduced current is due to intralayer and interlayer generation denoted by $I_{gen}$. $\vec{E_G}$ and $\vec{E_D}$ are the electric fields due to back gate and source-drain bias, respectively. The relative thickness of the arrows indicates the magnitude of the currents.

In contrast to pn junctions fabricated using inorganic doped semiconductors like Si, the forward bias current transport in vdW pn heterojunctions is dominated by tunnelling-based trap assisted (Shockley-Read-Hall, SRH) or direct (Langevin) recombination of majority carriers across the heterointerface.[6,47] The current transport due to Fowler-Nordheim tunnelling (FNT) through a triangular potential barrier is modelled by the equation:[48,49] $I \propto V^2 \exp\left(\frac{-4d\sqrt{2m^*\phi^3}}{3\hbar qV}\right)$, where $d$ is the thickness of the barrier, $m^*$ is the effective mass of the charge carriers, $\hbar$ is the reduced Planck's constant, and $\phi$ is the height of the tunnel barrier. Hence, the presence of tunnelling mediated transport can be confirmed by a straight line with negative slope in the $\ln(I_D/V_D^2)$ against $1/V_D$ plot. Figure 4b shows such a trend for forward bias $V_D$ varying from 0.12 V to 0.4 V at different $V_G$. The triangular barrier required for FNT could be at the heterointerface or at the contacts. For $V_G < -20$ V, near the $V_{Th}$ of WSe$_2$, $I_F$ is high, implying WSe$_2$ dominated forward bias transport.

Transfer characteristics across the heterostructure are shown in Figure 4c for $V_D = 1$ V and $V_D = -1$ V. The forward bias transfer curve exhibits prominent n and p branches owing to contributions from both materials. The $V_G$ dependent sheet carrier density of majority holes in WSe$_2$ is represented by $p_W$ and majority electrons in ReS$_2$ by $n_R$. For $V_G$ near the $V_{Th}$ in WSe$_2$ and ReS$_2$ FETs, the $I_D$ is approximated by the respective carrier concentrations ($n$) using the contact resistance ($R_C$) corrected drift current relation:[50,51] $I_D = qn\mu(V_D - I_D R_C)(W/L)$. The $p_W$ peaks at large negative $V_G$ (~ $5.3 \times 10^{12}$ cm$^{-2}$) and drops to its intrinsic value[6,52] ($p_{W0}$ ~ $3 \times 10^9$ cm$^{-2}$) below $V_{Th}$. On the other hand, $n_R$ is extracted to be ~ $1.8 \times 10^{13}$ cm$^{-2}$ at high positive $V_G$ but it does not display considerable variation with $V_G$. Further, $V_G$ dependent trends in the calculated values of $n_R$ and $p_W$ are shown in the inset of Figure 4c.

Being in close proximity to SiO$_2$, the Fermi level in thin WSe$_2$ can be effectively modulated by the gate bias. The electron affinities ($\chi$) of WSe$_2$ and ReS$_2$ are considered to be 4 eV[48] and 4.5 eV,[12] respectively. Figure 4d (i) shows the energy band diagrams highlighting the positions of the Fermi levels in WSe$_2$ ($E_{F\_W}$) and ReS$_2$ ($E_{F\_R}$) for $V_G = -40$ V. Based on the transfer characteristics of the individual FETs, $E_{F\_W}$ is assumed to be near the valence band and $E_{F\_R}$ to be near the conduction band. When forward biased, a positive voltage is applied to WSe$_2$ and ReS$_2$ is kept grounded. Now, as described in Figure 4d (ii), for $V_G = -40$ V, there is an accumulation of majority charge carriers- electrons from ReS$_2$ and holes from WSe$_2$ at the heterointerface which can undergo interlayer recombination to yield a large current component, $I_{rec} \propto n_R p_W$, marked as region (1) in Figure 4c ($n_R$ and $p_W$ both being large). For the reverse bias case in Figure 4d (iii), since the direction of the source-drain electric field ($\overrightarrow{E_D}$) is now reversed, the current (marked as region (5) in Figure 4c) is due to generation of charge carriers in WSe$_2$ (intralayer) or across the heterojunction (interlayer) and much smaller than the forward bias case. Further insights on the charge transport across the heterostructure for the various regions marked in Figure 4c and also at $V_G$ = 40 V are provided in S10 of Supporting Information.

## Photodetection

Optoelectronic characteristics of the three WSe$_2$/ReS$_2$ diode configurations (FL/FL, FL/8L, and 2L/2L) were studied using a 532 nm laser focused through a 50× objective yielding a spot of diameter ~ 4 $\mu$m. The laser power ($P_{laser}$) was varied from 1 nW to 0.5 mW. Unless otherwise stated, all $I_D$ –$V_D$ characteristics were recorded with an applied back gate voltage of −20 V for the FL/FL WSe$_2$/ReS$_2$ structure and −10 V for the two thinner configurations. For the FL/FL WSe$_2$/ReS$_2$ heterostructure, $I_D$– $V_D$ characteristics with increasing $P_{laser}$ are shown in Figure 5a. The current under illumination is denoted by $I_{light}$, from which the photocurrent ($I_{ph}$) is calculated as $I_{ph}$ = $I_{light}$ − $I_{dark}$. $I_{ph}$ depends on (i) generation of photoexcited electron-hole pairs (EHP) in the heterojunction region, (ii) efficient separation of the EHPs, and (iii) extraction of the carriers into the external circuit to contribute to the total current. Similar plots for the thinner heterostructures demonstrating lower photocurrents than the FL/FL diode for matched laser powers are shown in Figure S11 of Supporting Information. The high $I_{ph}$ obtained in the FL/FL WSe$_2$/ReS$_2$ device can be attributed to (i) FL flakes for higher optical absorption, (ii) near-direct bandgap of the WSe$_2$/ReS$_2$ heterojunction for enhanced optical generation, and (iii) high values of $\Delta E_C$ and $\Delta E_V$ which lead to efficient charge separation across the heterointerface. Photo-amplification is the ratio of $I_{light}$ to $I_{dark}$ at different $P_{laser}$ for a particular drain voltage. An enhanced photo-amplification exceeding 2×10$^6$ in the FL/FL WSe$_2$/ReS$_2$ heterostructure (Figure 5b) can enable excellent photodetection applications.

Responsivity ($R = I_{ph}/P_{laser}$) is a key figure of merit for photodetectors. A high $R$ of 3 A/W at 1 nW of laser power, better than commercial Si photodetectors (for 532 nm)[10] can be attributed to the large $I_{ph}$ generated in the overlap region owing to the high absorption coefficients of the thick WSe$_2$ and ReS$_2$ flakes.[53] Consequently, the FL/8L and 2L/2L WSe$_2$/ReS$_2$ heterostructures demonstrate lower $R$ that decreases with decreasing flake thicknesses. Absorption spectrum of a thick ReS$_2$ flake on sapphire substrate for small angle of incidence is shown in the Supporting Information, Figure S12. The variation of $R$ with $P_{laser}$ for the different heterostructures is shown in Figure 5c. A power dependence of $R \propto P_{laser}^{-0.42}$ for $V_D$ = 0 V is obtained by fitting the data. Detectivity ($D^*$) quantifies the photodetector's ability to measure very weak signals and can be calculated from the $R$ and $I_{dark}$ at a particular $V_D$ and $P_{laser}$ by the relation $D^* = \frac{R\sqrt{A}}{\sqrt{2qI_{dark}}}$. Here, $A$ is the area of the device. A $D^*$ value of $3 \times 10^{12}$ Jones at $V_D$ = 0 V and 1 nW illumination results in a noise equivalent power $\left(NEP = \frac{\sqrt{A}}{D^*}\right)$ of $2 \times 10^{-16}$ WHz$^{-1/2}$.

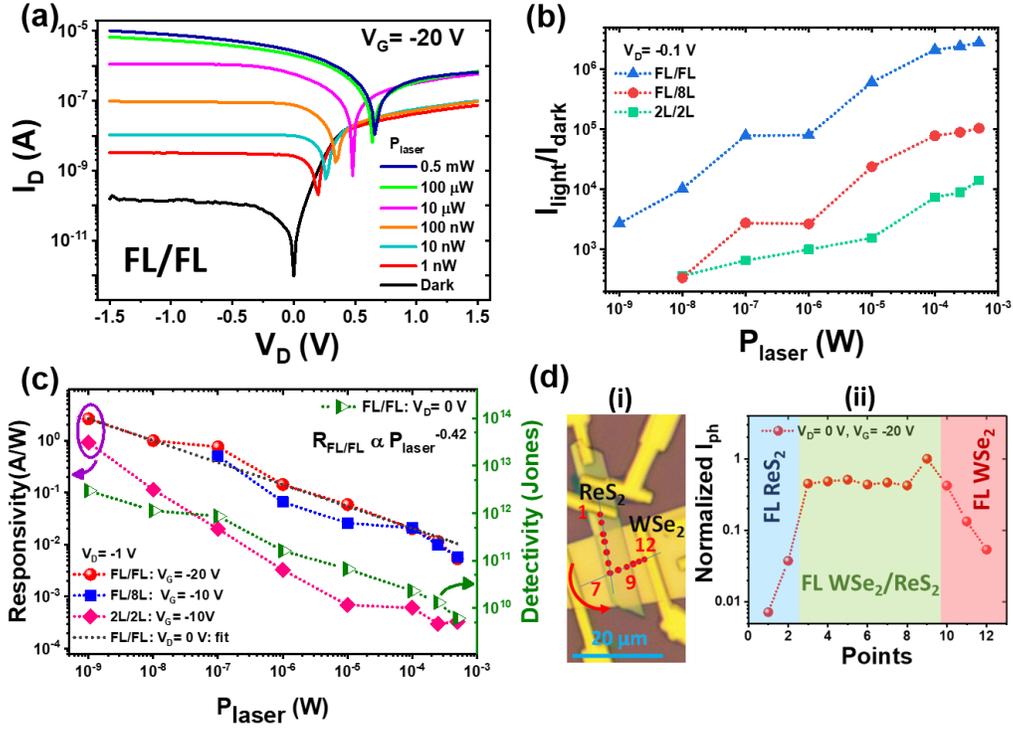

Figure 5: Photoresponse characteristics of WSe$_2$/ReS$_2$ heterojunctions. (a) I−V characteristics under 532 nm laser illumination at various incident powers for FL/FL heterostructure. (b) Photo-amplification with laser power for WSe$_2$/ReS$_2$ heterojunctions having different thicknesses of constituent flakes at a low $V_D$ of −0.1 V. (c) Responsivity for the different heterostructures, power law fit at $V_D$ = 0 V, and detectivity (of FL/FL structure) at various laser powers. (d) Linescan showing region-dependent normalized photocurrent at $V_D$ = 0 V, $V_G$ = −20 V. The laser beam was converged to a spot size of 1 μm diameter and focused on the points marked in red in (i). The photoresponse from the different regions: only FL WSe$_2$, FL/FL WSe$_2$/ReS$_2$ heterojunction, and only FL ReS$_2$, is shown in (ii).

The variation of $I_{ph}$ with $V_G$ under reverse bias ($V_D$ = −1 V) for the FL/FL heterostructure at 100 nW and 1 μW incident illumination is discussed in S13 of Supporting Information. As expected, the photocurrent increases with increasing illumination intensity and is significantly larger than the dark transfer characteristics. $I_{ph}$ vs $V_G$ for the forward biased diode and the transport mechanisms are included in Figure S14 of Supporting Information. To map the origin of the photoresponse, the laser was selectively incident on multiple points (1 to 12) in three distinct regions of the FL/FL heterojunction: (1) ReS$_2$ only, (2) WSe$_2$/ReS$_2$ overlap and, (3) WSe$_2$ only, as marked in Figure 4f (i). Normalized $I_{ph}$ for $V_D$ = 0 V, $V_G$ = −20 V is plotted in (ii) showing that the maximum current due to photogeneration is from the overlap region. In comparison to the individual WSe$_2$ and ReS$_2$ regions, the photocurrent shows 10-to-100 times increase in the overlap region, markedly higher than in BP/MoS$_2$,[1] WSe$_2$/SnS$_2$,[54] WSe$_2$/MoSe$_2$,[55] and MoS$_2$/GaSe[53] heterojunctions. This leads to a similar enhancement in $R$ at the heterointerface in comparison to regions 1 and 3.

## Photovoltaic effect and IR Photoresponse

A distinct photovoltaic effect is seen in the $I–V$ characteristics in Figure 5a. Figure 6a shows the fourth quadrant (photovoltaic operating region) plot for increasing $P_{laser}$ indicating the $V_{OC}$ and $I_{SC}$. In organic solar cells the difference between the highest occupied molecular orbital (HOMO) of the donor and the lowest unoccupied molecular orbital (LUMO) of the acceptor limits the maximum $V_{OC}$.[30] Here, the difference between the conduction band minimum of FL ReS$_2$ and the valence band maximum of FL WSe$_2$ at the heterointerface (0.7 eV) being larger than several other reported vdW heterostructures, a relatively higher $V_{OC}$ is predicted.[10,30] As expected, for the FL/FL heterostructure, a high $V_{OC}$ value of 0.64 V was obtained for $P_{laser}$ = 0.5 mW. The corresponding $I_{SC}$ was 2.6 μA. Slope of the $V_{OC}$ vs $P_{laser}$ plot (described in Figure S15 of Supporting Information) gives an ideality factor of ∼1.5 implying that both SRH and Langevin recombination processes influence the performance.

A modified Shockley diode equation[56]: $I(V) = I_{ph}\left\{\exp\left[\frac{q(V-V_{OC})}{k_BT(1+\alpha)}\right] - 1\right\}$, is used to fit the $I–V$ characteristics in the photovoltaic operation region. Here, $\alpha$ is a dimensionless quantity which depends on charge recombination and extraction.[56,57] An $\alpha$ value of ∼ 2, for 10 nW and 100 nW laser illumination indicates good carrier extraction at the contacts. Fits to the $I_D–V_D$ data for low illumination powers using the modified equation are shown in Figure 5b.

Power generated ($P_{elec}$) by the FL/FL photovoltaic device as a function of $V_D$ for various illumination powers is shown in Figure 6c. The output power reaches 350 nW for $P_{laser}$ = 0.5 mW, making it one of the highest values reported among vdW heterojunctions. The maximum electric power generated by the FL/8L and 2L/2L WSe$_2$/ReS$_2$ heterostructures, significantly (5× and 80×, respectively) lower than the FL/FL structure, is also shown in Figure 6c. Figure S16 of Supporting Information shows the variation in $I_{SC}$ and $V_{OC}$ with varying gate voltage and laser power. Similar to the transfer characteristics under illumination (S13 in Supporting Information), $I_{SC}$ increases with laser power and is higher for negative gate voltages. $V_{OC}$ is determined by the electron (ReS$_2$) and hole (WSe$_2$) quasi Fermi level difference upon illumination. In the FL/FL device, the Fermi level of ReS$_2$ stays near the conduction band and that of WSe$_2$ varies from the valence band edge upto midgap with back gate voltage. $V_{OC}$ shows good tunability with $V_G$, with the maximum for $V_G \leq -30$ V when the Fermi levels of WSe$_2$ and ReS$_2$ are near their respective valence and conduction band edges, thereby maximizing $E_{Fn}$ - $E_{Fp}$. Similarly, the minimum value for each incident power occurs for $V_G$ below (more positive) $V_{Th}$ of WSe$_2$. Further, the $V_{OC}$ depends on the dark current as well as the light induced photocurrent in the pn junction. Owing to the reduced thickness and lower absorption coefficients of the constituent flakes in the thinner heterostructures, the $V_{OC}$ is lower than the FL/FL structure as shown in Figure S17 of Supporting Information. As seen from Figure 6d, a maximum fill factor ($FF = P_{elec,\ max}/(V_{OC}I_{SC})$) of 0.4 and a maximum power conversion efficiency ($PCE = V_{OC}I_{SC}FF/P_{in}$) of 0.2 % are obtained. The FF and PCE of the thinner (2L/2L) heterostructure are compared to the thicker (FL/FL) device in Figure 6d. External quantum efficiency ($EQE = \frac{I_{ph}/q}{P_{laser}/h\nu} \times 100\%$) is ~600 % at 1 nW laser power. It should be noted that these values of $EQE$ have been calculated from corresponding values of $R$; the experimental values are typically found to be lower.[25]

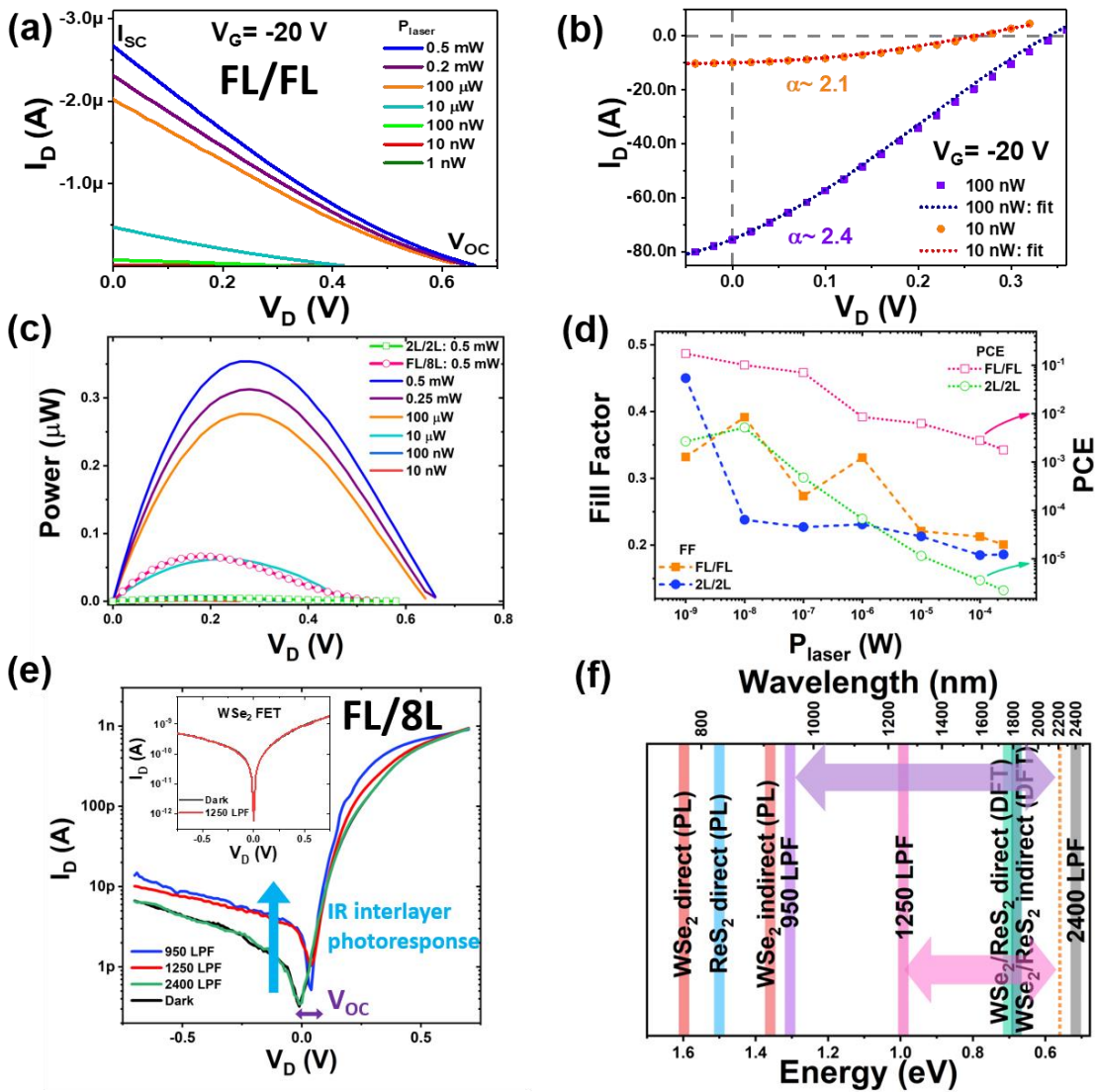

Figure 6: Photovoltaic effect and interlayer bandgap photoresponse in $WSe_2/ReS_2$ heterojunction. (a) Fourth quadrant plot showing $I_{SC}$ and $V_{OC}$ for varying laser power, and (b) fitting of photovoltaic data using the modified Shockley diode equation[56] for the FL/FL heterostructure. (c) Electrical power generated by the FL/FL device as a function of incident laser power. The maximum output electric power obtained for the FL/8L and 2L/2L heterostructures are also marked. (d) Fill factor and PCE at various laser powers for the FL/FL and 2L/2L heterostructures. (e) Output characteristics across the $WSe_2/ReS_2$ heterojunction under dark and upon IR illumination tuned using long pass filters. Photovoltaic effect and interlayer IR photoresponse are seen in the I-V characteristics. Inset shows that a $WSe_2$ FET does not have any response to the IR radiation above 1250 nm. (f) Plot depicting the energy range involved in the measurements.

The strong PL quenching and enhanced photocurrent generation from the heterojunction overlap region are in good agreement with DFT calculations that predict a type-II band alignment at the heterointerface. The bandgaps of FL WSe$_2$ and ReS$_2$ are 1.35 eV and 1.50 eV as per our PL measurements (S4 of Supporting Information). The 532 nm (2.330 eV) laser excitation can lead to intralayer as well as interlayer generation of EHPs in the individual WSe$_2$ and ReS$_2$ layers and across the heterointerface, respectively. Therefore, to demonstrate photogeneration exclusively across the interlayer bandgap, the FL/8L heterostructure was irradiated using an IR source and various long pass filters (LPF) were used to selectively block wavelengths below the cut-on wavelength of the LPF. For instance, a 1250 LPF would only allow wavelengths beyond 1250 nm to pass through it. Figure 6e shows output characteristics of the diode under dark and on exposure to IR radiation tuned using three LPFs: 950 nm (1.305 eV), 1250nm (0.992 eV), and 2400 nm (0.516 eV). Photocurrent is obtained when 950 nm and 1250 nm LPFs are used, which correspond to energies significantly lower than the bandgaps of the individual materials and higher than the DFT predicted value for the interlayer bandgap. A clear photovoltaic effect is also seen in the $I - V$ characteristics. Furthermore, the use of 2400 nm LPF yields negligible change from the dark characteristics. This observation rules out the contribution of sub-bandgap traps (if any) in the photocurrent generation across the interlayer bandgap.[58,59] Absence of photoresponse in an individual $WSe_2$ FET (inset in Figure 6e) when subjected to IR radiation under the same set-up reinforces the sub-bandgap (type-II) interlayer photogeneration across the heterointerface. A schematic depicting the energy range involved in the optical measurements in comparison to the DFT interlayer bandgap values and the bandgaps of individual $WSe_2$ and $ReS_2$ layers is shown in Figure 6f for better visualization.

## Photoswitching and Benchmarking

Transient photoresponse of the photodetector is a crucial measure for its use in high-speed applications. The three $WSe_2/ReS_2$ heterojunctions were exposed to laser illumination which was pulsed at frequencies ranging from 100 Hz to 50 kHz. The modulation of the photocurrent was recorded using a digital oscilloscope coupled to a low-noise amplifier. The rise time, $\tau_r$ (fall time, $\tau_f$) is calculated between 10% (90%) to 90% (10%) of the photocurrent. Figure 7a shows the photocurrent for the FL/FL heterodiode under 100 $\mu$W power switched at 10 kHz frequency, with both $V_D$ and $V_G$ kept at 0 V. This demonstrates the self-driven nature of the photodetection, owing to the photovoltaic effect, indicating good potential for ultra low-power applications. The $\tau_r$ and $\tau_f$ of ~ 5 $\mu$s make the FL/FL structure one of the fastest reported vdW heterojunctions. Figure 7b compares the photoswitching response of all three heterostructures at 10 kHz. The thinner heterostructures also exhibit fast switching, but slower than the FL/FL structure. The ultrafast response time can be partly attributed to the large band offsets ($\Delta E_C$, $\Delta E_V$) which enable prompt separation of the photogenerated charge carriers towards the contacts.[6] Variation of $\tau_r$ and $R$ with the switching frequency is shown in Figure 7c; indicating a possible trade-off between the two quantities as shown in other reports.[28] It may be noted that 5 $\mu$s is the limit of the measurement setup used in this work. For the FL/8L and 2L/2L heterostructures, the response times are ~ 24 $\mu$s and 28 $\mu$s, respectively. Stability is an essential parameter in the design space of vdW photodetectors since materials like BP, InSe are unstable[60] and can lead to degradation of device performance with time. Although capping layers can improve stability,[61] they can also lead to optical absorption loss and a lower external quantum efficiency. The choice of individual p and n materials in this work was also based on their air-stability without the need of a capping layer. As expected, the heterostructure shows good stability and a comparison of photoswitching response over a span of three months is shown in Figure 7d.

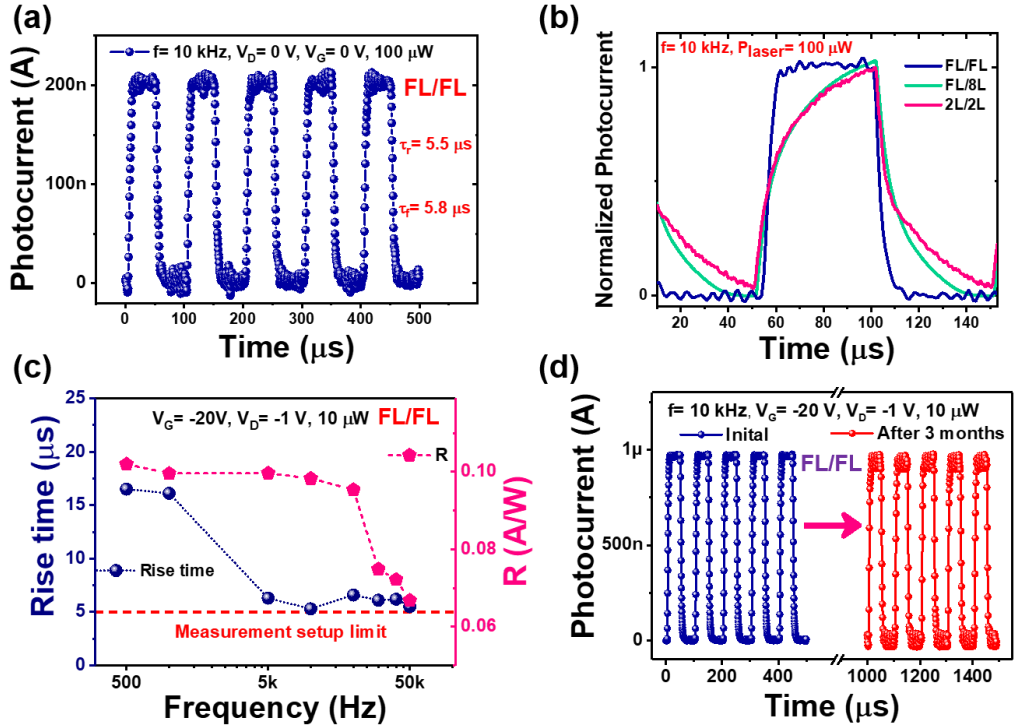

Figure 7: (a) Photoswitching at a frequency of 10 kHz with both $V_G$ and $V_D$ set at 0 V, showing self-powered operation for the FL/FL $WSe_2/ReS_2$ heterostructure. The rise and fall times are ∼ 5 $\mu$s. (b) Normalized photocurrent demonstrating fast switching response at 10 kHz for heterostructures of different thicknesses. (c) Variation of rise time and responsivity of the FL/FL device with switching frequency of the laser at 10 $\mu W$ power. (d) Photoswitching at 10 kHz and 10 $\mu W$ laser power before and after three months of ambient exposure demonstrating air stability of the FL/FL heterostructure.

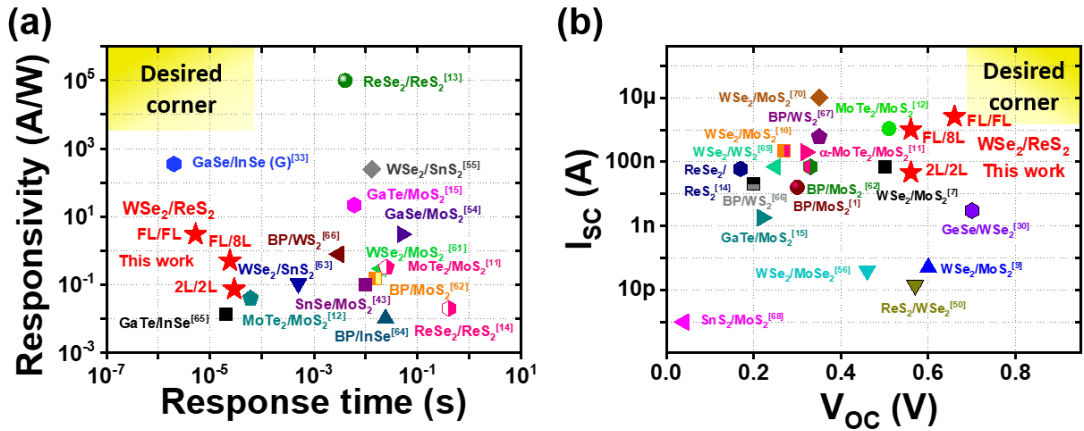

Figure 8: (a) Comparison of responsivity and response time of various vdW pn heterojunctions. This work demonstrates one of the fastest response times along with high responsivity. (b) Benchmarking of $I_{SC}$ and $V_{OC}$ of vdW pn heterojunctions with the current work showing excellent values of both parameters. $R$, $\tau_r$, $I_{SC}$, and $V_{OC}$ values for heterostructures of different thicknesses reported in this work are marked.

In order to benchmark the performance of our devices with published literature, a comparison of critical reported photodetector parameters, $R$ and $\tau_r$, are plotted in Figure 8a. Ideally, fast response time and high $R$ are desirable for photodetection applications (top-left corner of the plot). The high value of $R$ (3 A/W) and the ultrafast response time (5 $\mu$s) of the FL/FL WSe$_2$/ReS$_2$ heterostructure indicate significantly better performance than previously reported vdW heterostructure photodetectors.[10–12,29,32,42,54,62–67] To compare the photovoltaic performance, $I_{SC}$ is plotted against $V_{OC}$ for reported vdW heterostructures in Figure 8b. Excellent values of both, $V_{OC}$ (0.64 V) and $I_{SC}$ (2.6 $\mu$A), are obtained for the current work in comparison to other reports.[1,6,8,10,11,13,14,49,55,63,68–71]

## Conclusions

This work demonstrates that an appropriate choice of n- and p-type materials from the group-6 and -7 TMDs can lead to excellent photodetection and photovoltaic performance. DFT calculations for the heterojunction composed of few-layer flakes of ReS$_2$ stacked onto WSe$_2$ predict type-II band alignment with a near-direct interlayer bandgap at the heterointerface which is experimentally supported by PL and IR photocurrent measurements. Excellent electrical characteristics are demonstrated with gate voltage tunable current rectification reaching $10^5$. Dark and illuminated diode transfer characteristics have been explained based on interlayer recombination and generation processes. In terms of photodetection, one of the fastest (5 $\mu$s) high responsivity (3 A/W) transient photoswitching and the largest enhancement (10-100 ×) in photocurrent generation and responsivity at the overlap region are reported for the WSe$_2$/ReS$_2$ pn heterostructure owing to the type-II near-direct bandgap, high optical absorption, and efficient charge carrier separation due to large heterointerface band offsets. High $V_{OC}$ (0.64 V) resulting from gate tunability of WSe$_2$ and the 0.7 eV interlayer bandgap, large $I_{SC}$ (2.6 $\mu$A) and peak output power of 0.35 $\mu$W make this one of the best reported heterostructure photovoltaic devices. Further, flake thickness dependent optoelectronic performance justifies the choice of FL flakes with high absorption coefficients. Finally, long term ambient stability and ease of fabrication due to a single contact metal process without capping layers, coupled with the multi-functional and excellent optoelectronic performance, makes the FL WSe$_2$/ReS$_2$ pn heterostructure significantly advance the prospects of 2D layered semiconductors enabling next-generation optoelectronic applications.

# Methods

## Computational Methodology

In order to carry out electronic structure calculations and geometry optimizations, density functional theory was employed using the generalized gradient approximation (GGA) exchange correlation in conjunction with the Perdew-Burke-Ernzerhof (PBE) functional.[72] Moreover, as the norm-conserving pseudopotentials, the OPENMX (Open source package for Material eXplorer) code was used.[73,74] For FL structures, 9×9×3 k-points (in X, Y, and Z directions) along with the large density mesh cut off value of 200 Hartree were adopted. To account for the vdW interaction at different interfaces, the Grimme's dispersion correction (DFTD2) was utilized.[75]

## Device Fabrication and Characterization

The heterostructures were fabricated on an $SiO_2$ (285 nm)/ $p^+$ global back gate substrate. $ReS_2$ (HQ Graphene) and $WSe_2$ (SPI Supplies) were exfoliated onto PDMS stamps and selected flakes were transferred onto the substrate using a micromanipulator set-up. The heterostructure was cleaned in acetone to remove organic residues. The samples were spin coated with EL9/PMMA-A4 bilayer resist and electron beam lithography (Raith 150-Two) was followed by deposition of Cr (< 5 nm)/Pt (50 nm)/ Au (50 nm) stack (7-target sputtering system from AJA International) to define the contacts. Following metal lift-off in acetone, the heterostructure was annealed at 200 $^{\circ}$C to remove resist residues. Raman measurements were carried out on LabRAM HR800 (HORIBA Scientific) with a 1 $\mu$m spot diameter using a 532 nm laser. AFM measurements were performed using MFP-3D (Asylum Research Inc.). For PL measurements, the samples were irradiated with a 532 nm frequency doubled Nd-YAG laser via a 50× objective. The emitted luminescence was collected using a 0.5 m focal length Triax550 monochromator and detected using a Si-charge coupled device detector. For low temperature measurements, the samples were cooled in a liquid He flow cryostat with optical access.

Electrical measurements were carried out using a probe station connected to a Keysight B1500A parameter analyzer. For the photocurrent study, the metal contact pads on the substrate were wire-bonded onto a custom-made PCB and mounted on a home-built measurement box fixed on the stage of an Olympus BX63 microscope. The radiation from a 532 nm diode laser was incident on the heterostructure through a 50× objective (spot diameter of ~ 4 $\mu$m). Keysight B1500A was used for electrical measurements under illumination. For the photoswitching measurements, the laser beam was directly modulated by applying pulses from 100 Hz to 50 kHz using a function generator (Agilent 33220A). The current from the device was fed to an SR570 low noise current pre-amplifier and its output was recorded on an Agilent DSO6012A digital storage oscilloscope.

## Acknowledgement

A.V. thanks IITB-Monash Research Academy for the fellowship. D.S. acknowledges Department of Electrical Engineering, Indian Institute of Technology Bombay (IITB) for the Institute Post-Doctoral Fellowship. K.T. acknowledges Visvesvaraya PhD Scheme from the Ministry of Electronics and Information Technology (MeitY), Govt. of India. Authors acknowledge IITB Nanofabrication Facility (IITBNF) for usage of its facilities for device fabrication and characterization. N.V.M. acknowledges support from the Australian Research Council (CE170100039). This work was funded by the Department of Science and Technology, Govt. of India through the grant DST/SJF/ETA-01/2016-17. D.S. and K.T. contributed equally to the work.

## Supporting Information Available

SI is included.

# Supporting Information

# Near-direct bandgap WSe$_2$/ReS$_2$ type-II pn heterojunction for enhanced ultrafast photodetection and high-performance photovoltaics

Abin Varghese,[1,2,3] Dipankar Saha,[1] Kartikey Thakar,[1] Vishwas Jindal,[4] Sayantan Ghosh,[1] Nikhil Medhekar,[2] Sandip Ghosh,[4] and Saurabh Lodha[1*]

[1]Department of Electrical Engineering, Indian Institute of Technology Bombay, Mumbai 400076, India

[2]Department of Materials Science and Engineering, Monash University, Clayton, Victoria 3800, Australia

[3]IITB-Monash Research Academy, IIT Bombay, Mumbai 400076, India

[4]Department of Condensed Matter Physics and Materials Science, Tata Institute of Fundamental Research, Mumbai 400005, India

E-mail: slodha@ee.iitb.ac.in

## S1: Details of DFT calculations

DFT calculations were performed using the software package 'QuantumATK'.[1] For the electronic structure calculations and geometry optimizations, the generalized gradient approximation (GGA) was used as an exchange correlation along with the Perdew-Burke-Ernzerhof (PBE) functional.[2] Apart from that, as the norm-conserving pseudopotentials, "Open source package for Material eXplorer" (OPENMX) code was employed.[3,4] The basis set for W, Re, S, and Se atoms were taken as 's3p2d1', 's3p2d1', 's2p2d1', and 's2p2d1', respectively. The k-points in the Monkhorst-Pack grid were set as 9 × 9 × 1 (in X, Y, and Z directions) for calculating electronic structure of the unit cells, whereas for few-layer (FL) structures 9 × 9 × 3 k-points were adopted. Besides, the density mesh cut off was selected to a large value of 200 Hartree. Considering the FL structures, Grimme's dispersion correction DFTD2 was used to apprehend the van der Waals (vdW) interaction at the different interfaces.[5] Furthermore, to avoid any spurious interaction between periodic images, sufficient vacuum was incorporated along the perpendicular (Z) direction.

In order to optimize the geometries of the unit cells as well as FL $WSe_2/ReS_2$ heterostructures, the LBFGS (Limited-memory Broyden Fletcher Goldfarb Shanno) algorithm was utilized. Before performing these electronic structure calculations, the structures were optimized with a force tolerance value of 0.01 eV/ Å (along with a stress tolerance of 0.001 eV/Å$^3$).

Figure S1 shows (a, b) the fully relaxed unit cells and (c, d) the bandstructures of monolayer $WSe_2$ and $ReS_2$. The optimized lattice constants for the triclinic $ReS_2$ unit cell were obtained as a= 6.55 Å, b= 6.46 Å, and c= 12.77 Å (with α= 104.05°, β= 91.76°, and γ= 118.86°).[6,7] For monolayer $ReS_2$, the direct band gap located at Γ point is ~1.476 eV. However, strictly speaking, the valence band maximum is not at Γ point (rather it lies in between Γ and Y points). Hence, monolayer $ReS_2$ exhibits a near-direct band gap of 1.471 eV. On the other hand, monolayer $WSe_2$ is a direct bandgap (~1.6 eV) semiconductor. The optimized lattice constants are a= b= 3.32 Å for the hexagonal $WSe_2$ unit cell.[11]

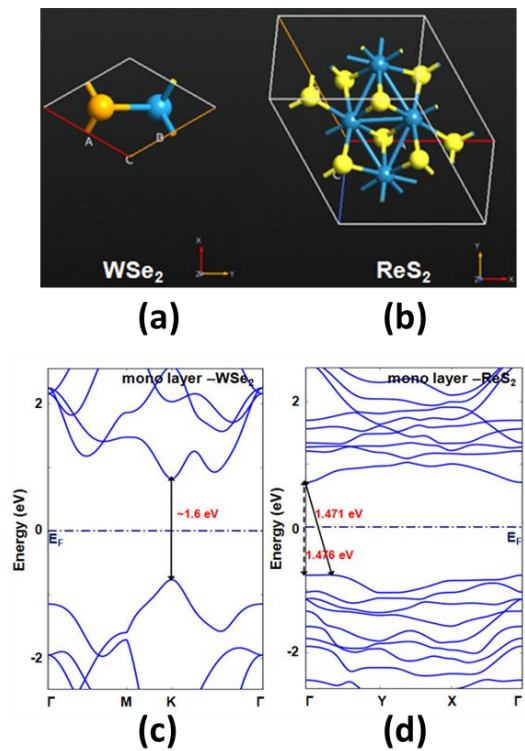

**Figure S1.1:** Geometry optimized unit cells of (a) $WSe_2$ and (b) $ReS_2$, portrayed in the X-Y plane. Considering the X-Y plane view of the $WSe_2$ unit cell, it is important to note that the upper 'Se' atom is exactly on top of the lower one. Electronic bandstructures of monolayer $WSe_2$ and monolayer $ReS_2$ are shown in (c) and (d), respectively. $E_F$ denotes the position of the Fermi level.

Tables T1 and T2 illustrate the change in bandgaps with the increase in the number of layers of $WSe_2$ and $ReS_2$, respectively. Layer dependent DFT calculated bandgaps and the ones obtained via photoluminescence (PL) are in good agreement with the corresponding literature.

| | $WSe_2$ | | | | | | | | | | | |
|---|---|---|---|---|---|---|---|---|---|---|---|---|
| | This Work | | | Bhattacharyya et al.[8] | | | Kam et al.[9] | | | Zhao et al.[10] | | |
| No. of layers | Band gap (eV) | Method | Direct/ Indirect | Band gap (eV) | Method | Direct/ Indirect | Band gap (eV) | Method | Direct/ Indirect | Bandgap (eV) | Method | Direct/ Indirect |
| 1 | 1.59 1.65 | Computational (GGA PBE + DFTD2) + Experimental | D | 1.53 | Computational (PBE + vdW) | D | --- | Experimental | --- | --- | Experimental | --- |
| 2 | 1.46 1.53 | | I | 1.43 | | I | --- | | --- | 1.53 (K-Γ) | | I |
| 3 | 1.32 | | I | --- | | --- | --- | | --- | 1.45 (K-Γ) 1.42 (Λ-Γ) | | I |
| 4 | 1.24 | | I | --- | | --- | --- | | --- | 1.42 (K-Γ) 1.37 (Λ-Γ) | | I |
| 5 | 1.21 | | I | --- | | --- | --- | | --- | --- | | --- |
| 6 | 1.18 | | I | --- | | --- | --- | | --- | --- | | --- |
| 8 | 1.16 | | I | --- | | --- | --- | | --- | 1.37 (K-Γ) 1.28 (Λ-Γ) | | I |
| 10 | 1.15 | | I | --- | | --- | --- | | --- | --- | | --- |
| FL | 1.35 | | I | --- | | --- | --- | | --- | --- | | --- |
| Bulk | | | --- | 1.08 | | I | 1.20 | | I | --- | | --- |

**Table T1**: Comparison of the thickness dependent calculated (in green) bandgaps of $WSe_2$ with literature. The bandgaps obtained via photoluminescence (in red) measurements for flakes of different thickness are also compared with literature.

| No. of layers | This Work | | | He et al.[12] | | | Marzik et al.[13] | | | Tongay et al.[14] | | |
|---|---|---|---|---|---|---|---|---|---|---|---|---|
| | Band gap (eV) | Method | Direct/Indirect | Band gap (eV) | Method | Direct/Indirect | Bandgap (eV) | Method | Direct/Indirect | Bandgap (eV) | Method | Direct/Indirect |
| 1 | 1.476/ 1.471 | Computational (GGA PBE + DFTD2) + Experimental | D/I | 1.569/ 1.532 | (LDA PZ) | D/I | --- | Experimental | | 1.43 1.59 | Computational (GGA PBE) + Experimental | D/I |
| | | | | 1.431/ 1.424 | (GGA PBE) | D/I | --- | | | | | |
| 2 | 1.40 1.50 | | D | --- | Computational | | --- | | | 1.51 | | |
| 3 | 1.37 | | D | --- | | | --- | | | | | |
| 4 | 1.37 | | D | --- | | | --- | | | 1.50 | | |
| FL | 1.50 | | --- | --- | | | | | | | | |
| Bulk | --- | | --- | --- | | | 1.32 | | I | 1.35 | | D/I |

**Table T2**: Comparison of the thickness dependent calculated (in green) bandgaps of ReS$_2$ with literature. The bandgaps obtained via photoluminescence (in red) measurements for flakes of different thickness are also compared with literature.

Next, the atomistic models for various heterostructures were designed using the geometry optimized WSe$_2$ and ReS$_2$ unit cells. The composite heterointerfaces consist of *m* layers (*m*L) of WSe$_2$ and *n* layers (*n*L) of ReS$_2$ (where m= 3, 4, 5, and 6 and n= 2, 3, and 4) and are represented as *m*L/*n*L WSe$_2$/ReS$_2$. The direct and indirect bandgaps as well as the corresponding ΔE (energy difference between direct and indirect bandgap values) for 3L/2L WSe$_2$/ReS$_2$, 4L/2L WSe$_2$/ReS$_2$, 5L/3L WSe$_2$/ReS$_2$, and 6L/4L WSe$_2$/ReS$_2$ heterostructures are listed in Table T3. It can be observed that for 3L/2L WSe$_2$/ReS$_2$ heterostructure, difference between direct and indirect bandgaps is comparatively large (0.057 eV). Beyond 3L/2L WSe$_2$/ReS$_2$, as the number of layers of the constituent materials is increased, ΔE values remain close-to 0.02 eV. As ΔE did not change significantly in thicker heterostructures, the electronic properties of the 4L/2L WSe$_2$/ReS$_2$ heterointerface were emphasized in this work (Figure 1).

For the composite 4L/2L WSe$_2$/ReS$_2$ heterointerface, the mean absolute strain on 4 layer-WSe$_2$ (as well as on 2 layer-ReS$_2$) was restricted to a small value of 0.66%. Besides, the stacking patterns for 4 layer-WSe$_2$ and 2 layer-ReS$_2$ were found to be in good agreement with previous studies.[8,12] Next, the heterointerface was wrapped into a hexagonal supercell and fully relaxed with the aforementioned force and stress tolerance values to construct the final structure. The optimized lattice constants for the 4L/2L WSe$_2$/ReS$_2$ vdW heterostructure are a= b= 6.621 Å.

| WSe$_2$/ReS$_2$ heterostructure | 3L/2L | 4L/2L | 5L/3L | 6L/4L |
|---|---|---|---|---|
| Direct bandgap (eV) | 0.7365 | 0.7059 | 0.7402 | 0.7290 |
| Indirect bandgap (eV) | 0.6790 | 0.6827 | 0.7166 | 0.7115 |
| **ΔE (eV)** | **0.0575** | **0.0232** | **0.0236** | **0.0175** |

**Table T3**: Calculated direct and indirect bandgap values along with their corresponding difference (ΔE) for various FL WSe$_2$/ReS$_2$ heterostructures.

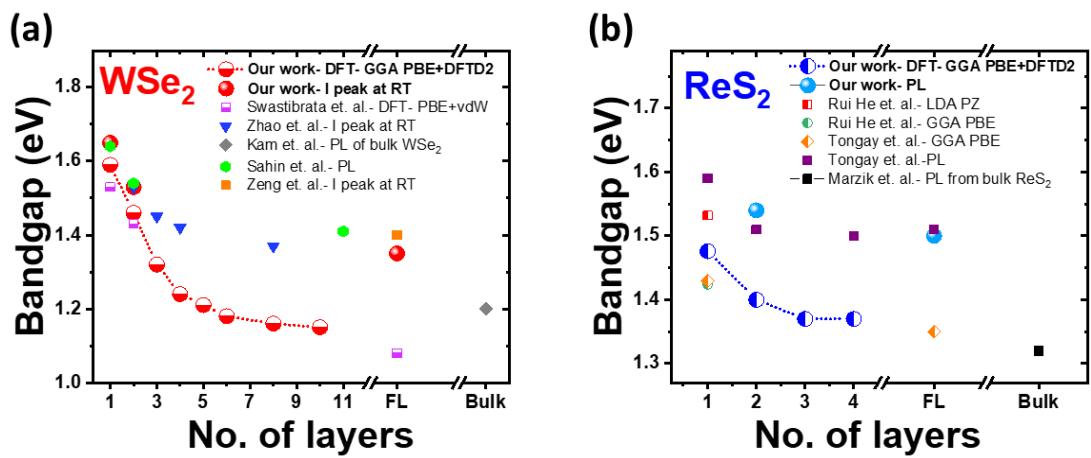

**Figure S1.2**: Comparison of calculated (DFT) and experimentally (PL) obtained bandgaps of (a) WSe$_2$ and (b) ReS$_2$ from this work and literature.

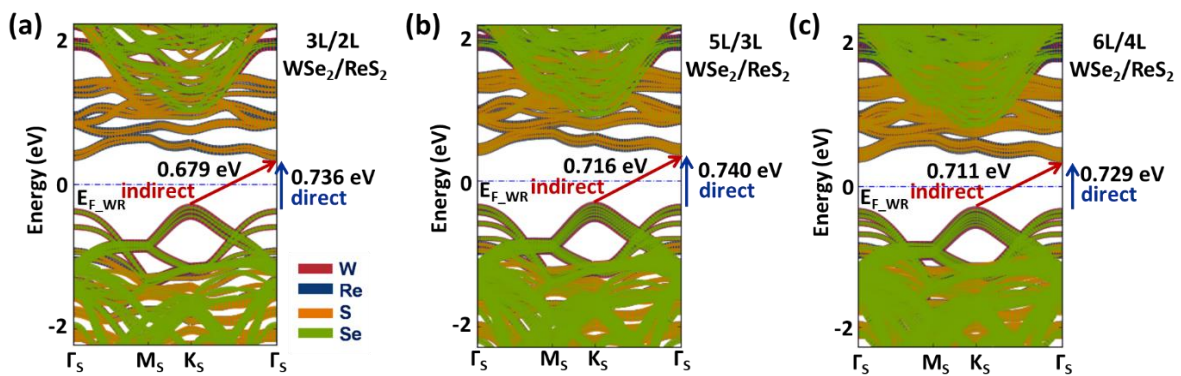

**Figure S1.3**: Projected bandstructures of various WSe$_2$/ReS$_2$ vdW heterostructures showing element-wise contributions. E$_{F\_WR}$ denotes the Fermi level position. The geometry optimized lattice constants and mean absolute strain values for various FL heterostructures were a= b= 6.61 Å and 0.66 % for 3L/2L WSe$_2$/ReS$_2$, a= b= 6.61 Å and 0.62 % for 5L/3L WSe$_2$/ReS$_2$, and a= b= 6.61 Å and 0.67 % for 6L/4L WSe$_2$/ReS$_2$.

## S2: Optical microscope image of FL/FL WSe$_2$/ReS$_2$ heterostructure

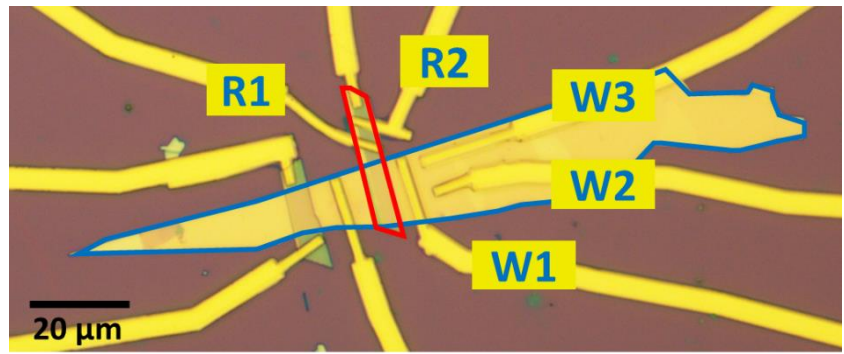

**Figure S2**: Optical microscope image of the FL/FL WSe$_2$/ReS$_2$ heterostructure presented in Figure 2. The WSe$_2$ flake is marked in blue and the ReS$_2$ flake in red.

Field effect transistors (FET) characteristics were studied for WSe$_2$ using the contacts W2 and W3 and for ReS$_2$ using the contacts R1 and R2. The diode characteristics were measured across the contacts R1 and W1.

## S3: AFM images of WSe$_2$/ReS$_2$ heterostructures

AFM images and linescans of heterostructures of different thicknesses. The thickness values for the WSe$_2$ and ReS$_2$ are consistent with literature. [14,15]

1. <u>FL/FL WSe$_2$/ReS$_2$ heterostructure</u>: A second FL/FL device besides the one reported in the main manuscript (Figure 2)

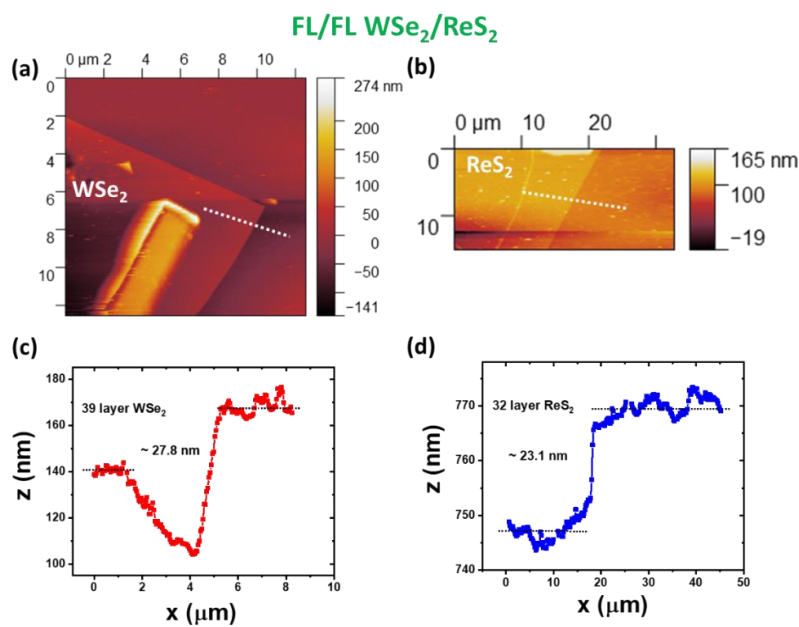

**Figure S3.1**: AFM images of (a) FL WSe$_2$ and (b) FL ReS$_2$ and the corresponding (c,d) linescans showing the thickness of the flakes.

2. FL/8L WSe$_2$/ReS$_2$ heterostructure: 65L/8L

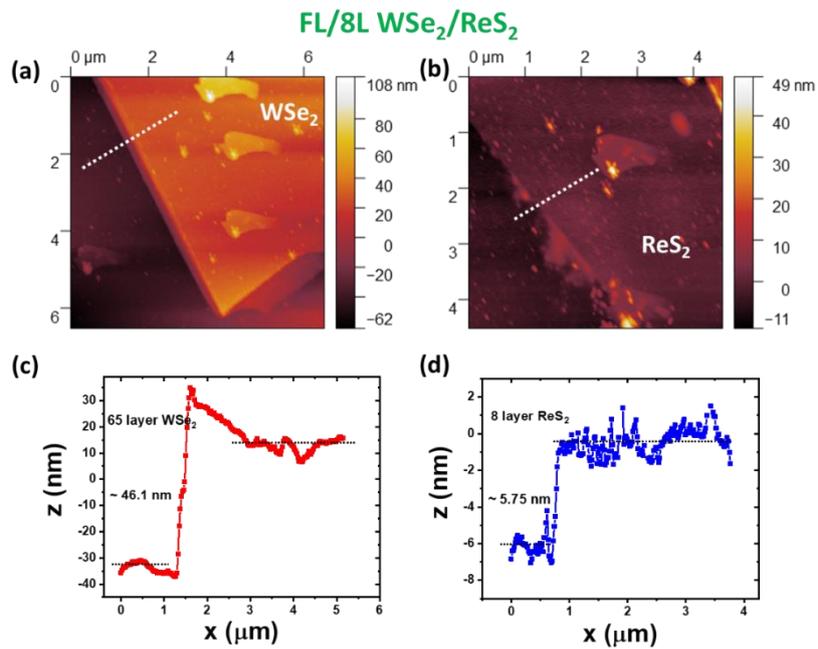

**Figure S3.2**: AFM images of (a) FL WSe$_2$ and (b) 8L ReS$_2$ and the corresponding (c,d) linescans showing the thickness of the flakes.

3. 2L/2L WSe$_2$/ReS$_2$ heterostructure

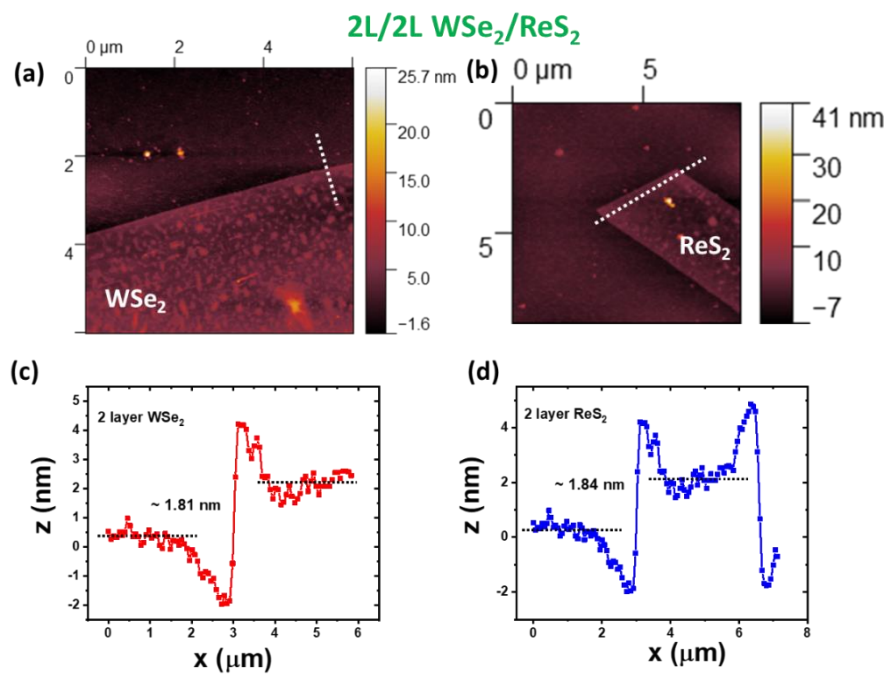

**Figure S3.3**: AFM images of (a) 2L WSe$_2$ and (b) 2L ReS$_2$ and the corresponding (c,d) linescans showing the thickness of the flakes

4. **S4: Photoluminescence studies on WSe$_2$/ReS$_2$ heterostructures**

Room temperature and low-temperature photoluminescence characterization was carried out to study the optical properties of the heterointerface. WSe$_2$/ReS$_2$ heterostructures with three distinct thickness configurations were considered:

1. Thick WSe$_2$ (65 layers)/ Thin ReS$_2$ (8 layers) → **FL/8L WSe$_2$/ReS$_2$**
2. Thin WSe$_2$ (2 layers)/ Thin ReS$_2$ (2 layers) → **2L/2L WSe$_2$/ReS$_2$**
3. Thick WSe$_2$ (32 layers)/ Thick ReS$_2$ (28 layers): Similar to the heterostructure described in the main manuscript → **FL/FL WSe$_2$/ReS$_2$**

The measured PL peak positions are as follows:

| Temp (K) | 65L WSe$_2$ | | 32L WSe$_2$ | | 2L WSe$_2$ | | 28L ReS$_2$ | 8L ReS$_2$ | 2L ReS$_2$ |
|---|---|---|---|---|---|---|---|---|---|
| Peaks | A | I | A | I | A | I | D | D | D |
| 300 | 1.59 | 1.35 | 1.59 | 1.35 | 1.60 | 1.53 | 1.50 | 1.50 | 1.54 |
| 25 | | | | | 1.69 | 1.53 | | | 1.6 |

Where, A is the transition from K of conduction band to K of valence band of WSe$_2$, I is the transition from the conduction band minima between Γ and K to Γ of WSe$_2$. Further, the D peak is due to the direct bandgap transition at Γ of ReS$_2$.

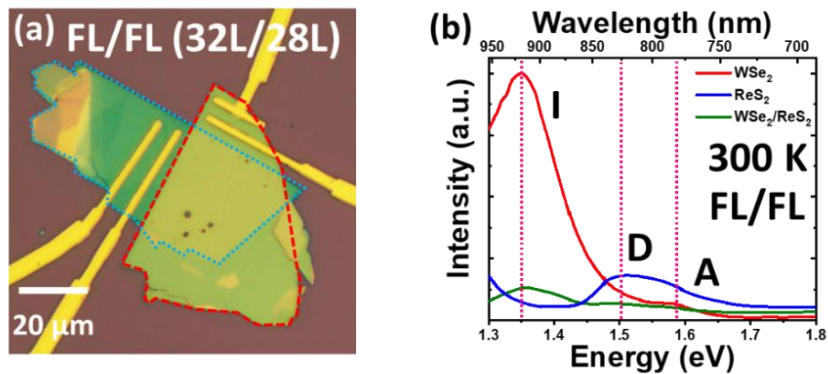

**Figure S4**: (a) Optical microscope image of FL/FL (32L/28L) WSe$_2$/ReS$_2$ heterostructure. (b) Room temperature PL peaks from individual materials and in WSe$_2$/ReS$_2$ heterostructure overlap region. There is a significant quenching of PL in the overlap region.

## S5: Transfer characteristics of individual WSe$_2$ and ReS$_2$ FETs

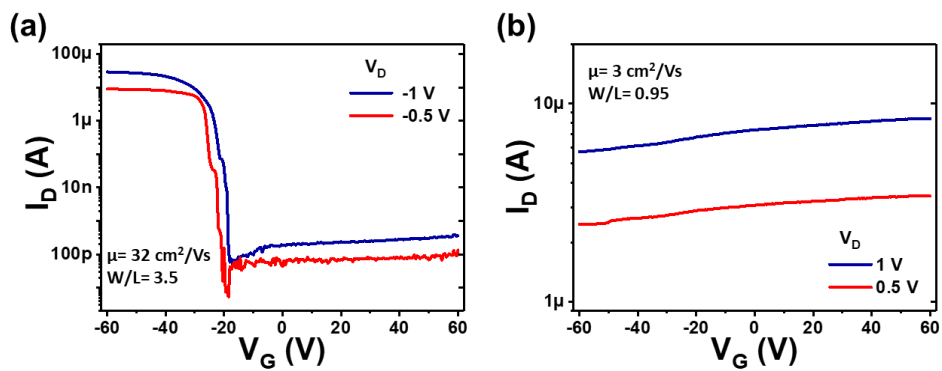

**Figure S5.1**: Transfer characteristics of (a) FL WSe$_2$ and (b) FL ReS$_2$.

Clear p- and n-type transport is observed in WSe$_2$ and ReS$_2$, respectively. The values of mobility (µ) and W/L are also marked.

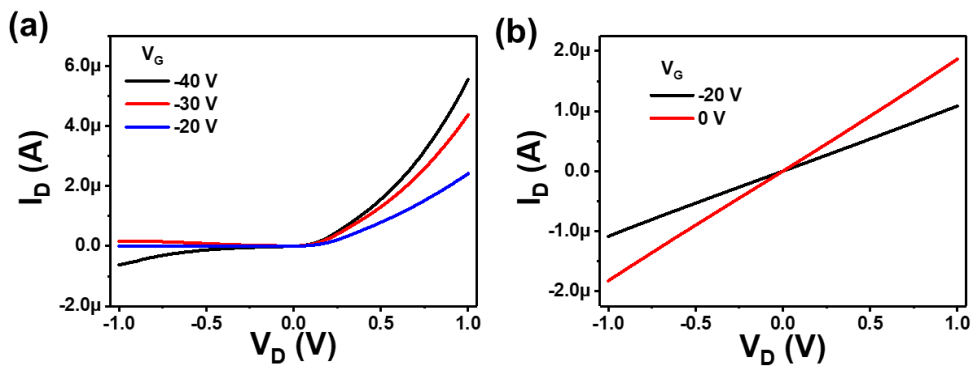

**Figure S5.2**: Output characteristics of (a) FL WSe$_2$ and (b) FL ReS$_2$.

The WSe$_2$ FET shows Schottky barrier dominated current transport. The ReS$_2$ FET displays ohmic characteristics.

## S6: High on/off ratio in thin ReS$_2$ FET

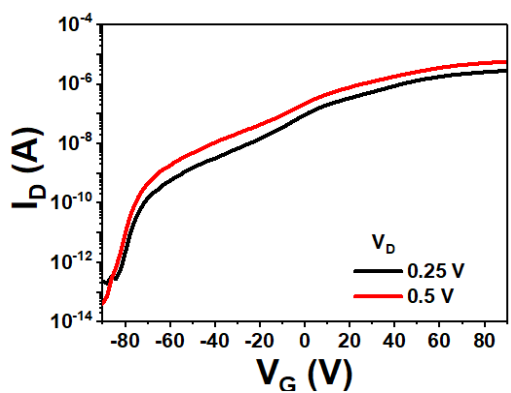

**Figure S6**: Transfer characteristics of thin ReS$_2$ FET.

In contrast to the transfer characteristics of the relatively thicker ReS$_2$ FET shown in Figure S5.1 (a), the transfer characteristics of trilayer ReS$_2$ (in S6) exhibits good modulation of the drain current with gate voltage. The current on/off ratio is $10^8$.

## S7: Pt contacts: without Cr adhesion layer

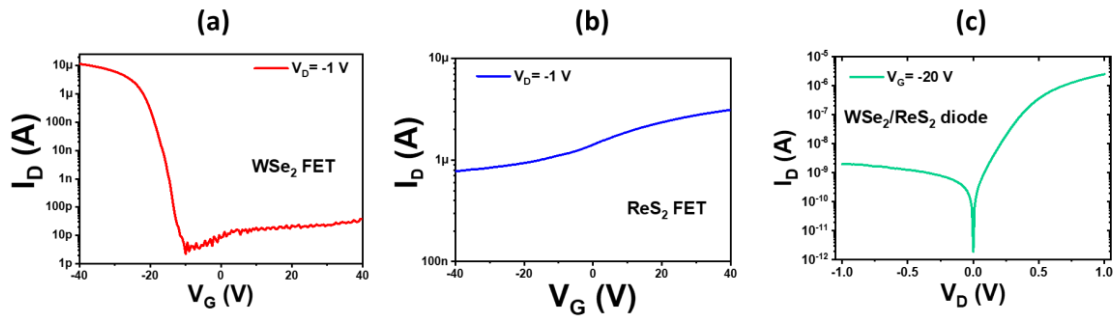

**Figure S7**: Transfer characteristics of (a) WSe$_2$ and (b) ReS$_2$ FET with Pt (50 nm)/Au (50 nm) contact. (c) I-V characteristics of the FL/FL WSe$_2$/ReS$_2$ heterojunction diode.

Pt is the choice of contact metal for our study. However, < 5 nm Cr is used as an adhesion layer to improve the sticking of Pt on Si/SiO$_2$ substrate. The use of such thin layers of Cr or Ti has been frequently employed during sputtering to improve the adhesion of metals like Au and Pt on Si/SiO$_2$ substrate. The transfer characteristics for the heterostructures without the thin Cr layer (above) are similar to those in Figure S5.

## S8: Gate tunable I-V characteristics of a FL/FL WSe$_2$/ReS$_2$ diode with a maximum rectification ratio of $10^5$

I-V characteristics of a similar FL/FL WSe$_2$/ReS$_2$ heterostructure are shown in (a). Gate tunable rectification ratio and ideality factor are plotted in (b). The maximum rectification obtained is $10^5$ for V$_G$= -20 V. The ideality factor is ~2 and the current transport is recombination dominated.

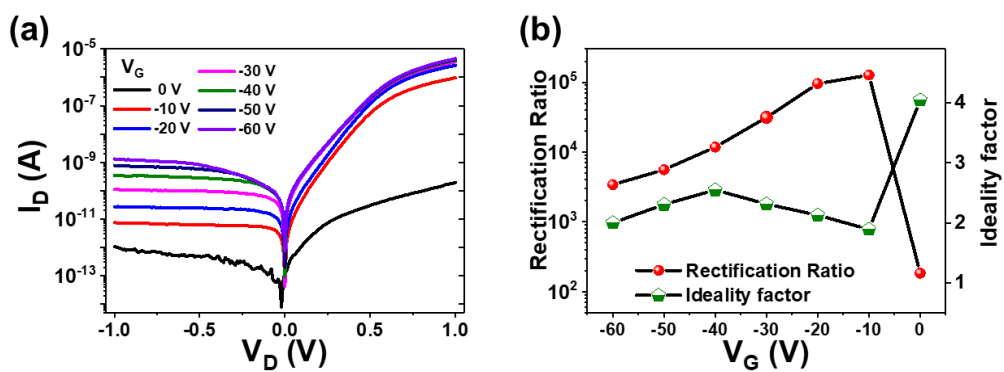

**Figure S8**: Gate-tunable (a) I-V characteristics and (b) rectification ratio and ideality factor of a similar FL/FL WSe$_2$/ReS$_2$ heterojunction diode.

## S9: Diode I-V characteristics for positive $V_G$

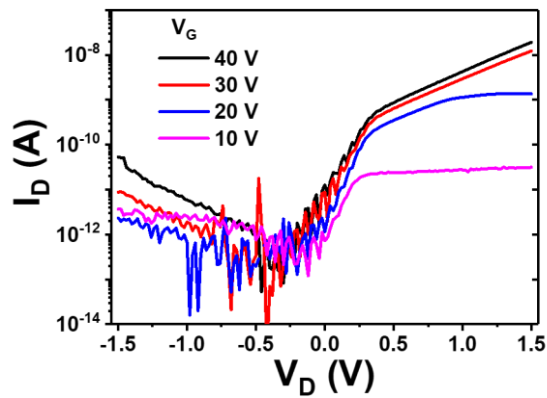

**Figure S9**: I-V characteristics of the FL/FL WSe$_2$/ReS$_2$ (Figure S2) heterojunction diode for positive gate voltage.

I-V characteristics of the FL/FL WSe$_2$/ReS$_2$ diode for positive gate voltages are shown. The forward currents are lower as compared to the negative gate voltage data described in the main text. The observed characteristics agree with the transfer curves of the heterostructure which demonstrates lower drain currents at positive gate bias.

## S10: Electrical transport across FL/FL WSe$_2$/ReS$_2$ heterojunction diode

Being in close proximity to SiO$_2$, the Fermi level in thin WSe$_2$ can be effectively modulated by the gate bias. The electron affinities ($\chi$) of WSe$_2$ and ReS$_2$ are considered to be 4 eV and 4.5 eV, respectively. Figure S10.2 shows the energy band diagrams highlighting the positions of the Fermi levels in WSe$_2$ ($E_{F\_W}$) and ReS$_2$ ($E_{F\_R}$) for $V_G$ = −40 V and $V_G$ = 40 V, respectively. Based on the transfer characteristics of the individual FETs, $E_{F\_W}$ is assumed to be near the valence band and $E_{F\_R}$ to be near the conduction band. On the other hand, at $V_G$ = 40 V, the WSe$_2$ FET being in the off state, $E_{F\_W}$ is taken to be near the intrinsic level, whereas $E_{F\_R}$ is mostly unchanged.

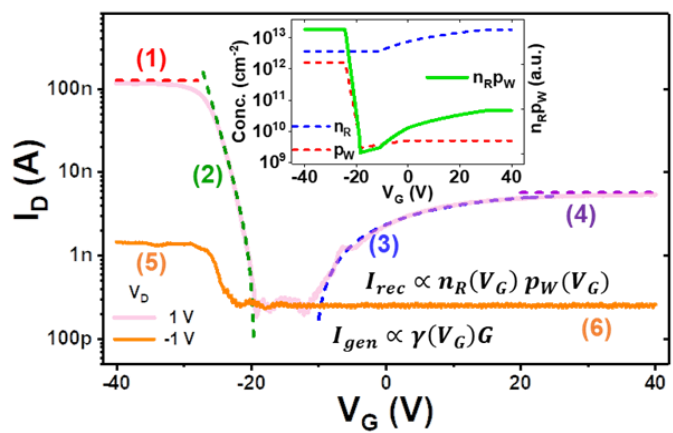

**Figure S10.1**: Transfer characteristics of the pn heterojunction for forward and reverse bias conditions with the various regions marked. Inset shows gate voltage dependent extracted hole sheet density in WSe$_2$ ($p_W$) and electron sheet density in ReS$_2$ ($n_R$) along with the product $n_R p_W$ that determines the interlayer recombination current.

When forward biased, a positive voltage is applied to WSe$_2$ and ReS$_2$ is kept grounded. Now, as described in Figure S10.2, for V$_G$ = −40 V, there is an accumulation of majority charge carriers- electrons from ReS$_2$ and holes from WSe$_2$ at the heterointerface which can undergo interlayer recombination to yield a large current component, $I_{rec} \propto n_R p_W$, marked as region (1) in Figure S10.1 (n$_R$ and p$_W$ both being large). Additionally, the transport of majority electrons (holes) from the conduction (valence) band of ReS$_2$ (WSe$_2$) to the conduction (valence) band of WSe$_2$ (ReS$_2$) contribute to the over-the-barrier current (I$_{maj}$), which is less pronounced compared to I$_{rec}$. It should be noted that V$_{Th}$ of the WSe$_2$ FET and of the p-branch of the forward bias transfer characteristic of the diode are similar. I$_D$ shows an exponential decay marked by region (2) in Figure S10.1, which corresponds to a sub-threshold current ($I_D \propto \exp\left(\frac{q(V_G - V_{Th})}{k_B T}\right)$) due to transport dominated by WSe$_2$.

Thereafter, for V$_G$ > V$_{Th}$ of WSe$_2$, region (3) shows an increasing current due to contribution from ReS$_2$. A polynomial function of the nature $I_D \propto A + B_1 x + B_2 x^2$ can be employed to fit the curve in region (3); the fit parameters are similar to those for the transfer characteristic of the ReS$_2$ FET. Finally, region (4) shows a nearly constant current defined by interlayer recombination, the magnitude diminished with respect to negative V$_G$ owing to the reduced p$_W$, since WSe$_2$ is now in the off state. The inset in Figure S10.1 shows the variation of n$_R$p$_W$ based on calculated carrier densities and shows good resemblance with the forward bias diode transfer curve.

For the reverse bias case in Figures S10.2 a(iii) and b(vi), since the direction of the source-drain electric field (**E$_D$**) is now reversed, the current is due to generation of charge carriers in WSe$_2$ (intralayer) or across the heterojunction (interlayer). The generation current (I$_{gen}$) takes the form, I$_{gen}$ ∝ γ(V$_G$)Gt$_{eff}$. Here, γ is the charge carrier extraction efficiency which depends on V$_G$ of the gated diode. G is the generation rate and t$_{eff}$ is a V$_D$ dependent length scale over which the generation processes take place. In addition to the electrostatic modulation of n$_R$ and p$_W$, V$_G$ also determines the effective barrier for the transport of electrons and holes across the heterojunction. Unlike n$_R$, p$_W$ shows good tunability with V$_G$. Hence the dependence of γ on p$_W$ is evident. The difference in reverse bias I$_D$ between regions (5) and (6) in Figure 3d is due to the change in γ. The magnitude of γ is much larger for negative V$_G$ due to significantly larger p$_W$ in WSe$_2$ at negative gate biases.

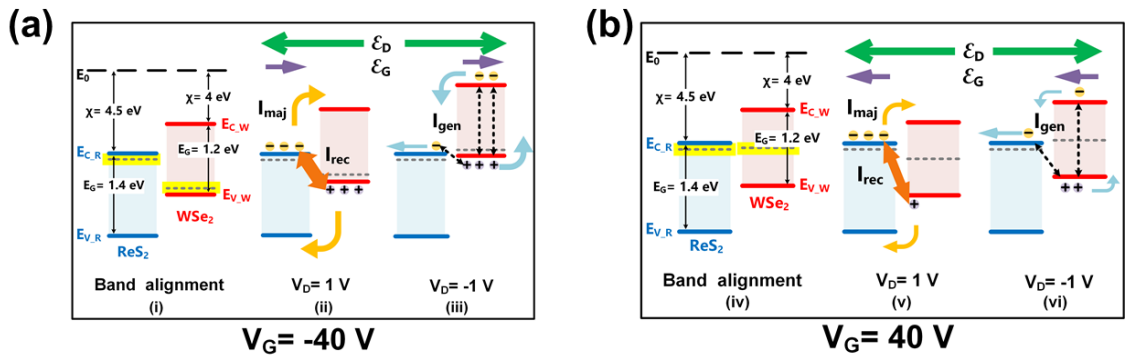

**Figure S10.2**: (a) Energy band diagrams for V$_G$ = -40 V: (i) band alignment in WSe$_2$ and ReS$_2$ with their respective Fermi levels defined by the gate voltage, (ii) under forward bias, the dominant charge transport is governed by interlayer recombination of the majority carriers giving rise to I$_{rec}$, and (iii) for reverse bias the reduced current is due to intralayer and interlayer generation denoted by I$_{gen}$. (b) At V$_G$ = 40 V, (iv) shows band alignment with the position of the Fermi levels, (v) under forward bias, I$_{rec}$ is lower than the V$_G$ = -40 V case due to lower p$_W$, and (vi) under reverse bias, a reduced (V$_G$) results in lower I$_{gen}$. **E$_G$** and **E$_D$** are the electric fields due to back gate and source-drain bias, respectively. The relative thickness of the arrows indicates the magnitude of the currents.

## S11: Photocurrent in thinner (FL/8L and 2L/2L) heterostructures

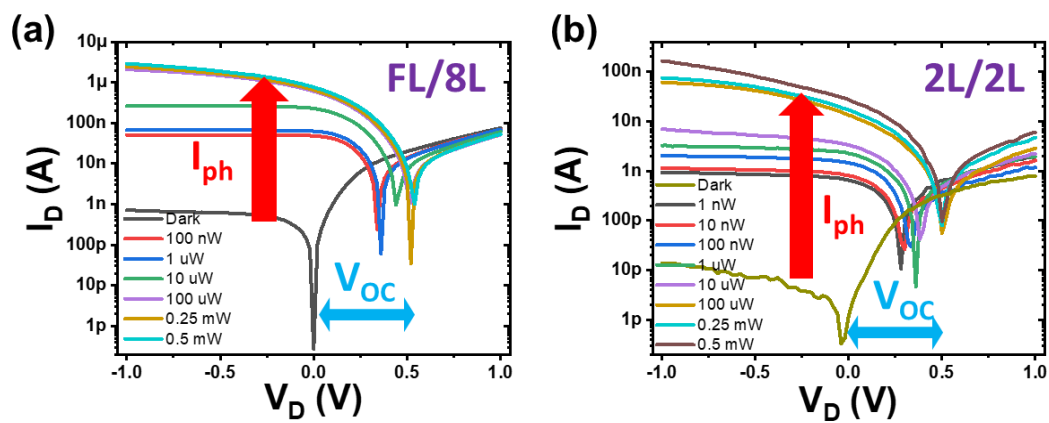

**Figure S11**: The I-V characteristics under 532 nm focused laser illumination are shown for (a) FL/8L and (b) 2L/2L heterostructure diodes. Photocurrent generation and $V_{OC}$ are highlighted.

The photocurrent is lower for the heterostructures composed of thinner flakes owing to the lower absorption of incident radiation and thereby, reduced generation of electron-hole pairs.

## S12: Absorption spectrum of thick ReS$_2$ flake

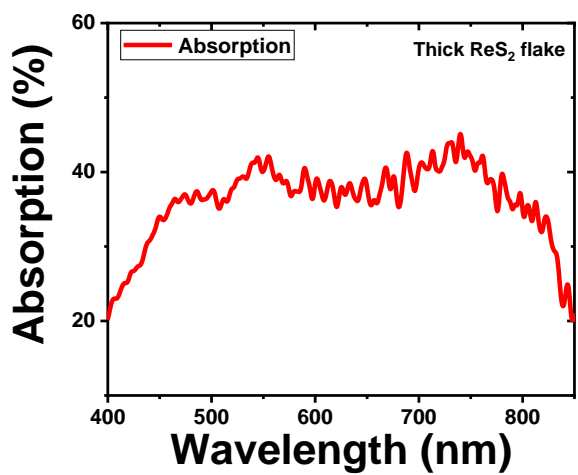

**Figure S12**: Absorption spectrum of thick ReS$_2$ flake on sapphire substrate measured for small angle of incidence ($15^0$) shows high optical absorption.

## S13: Reverse bias transfer characteristics under illumination for negative $V_G$ (FL/FL structure)

The generation current under reverse bias (Figure S12a) can be described as, $I_{gen} \propto \gamma(V_G)G_L t_{eff}$, where $G_L$ is the photogeneration rate due to the incident radiation. The dependence on γ (hence, $V_G$) is the same for illuminated and dark characteristics since illumination does not change the dependence of $p_W$ on $V_G$ that determines γ. Electric field due to the gate bias ($E_G$) at the vertical overlap region is parallel to $E_D$ for negative $V_G$ and is in the opposite direction for positive $V_G$ (Figure S12b). Hence, for negative $V_G$, the gate-induced and the photogenerated electrons (holes) in $ReS_2$ ($WSe_2$) are both directed towards the contacts which leads to a larger net current compared to positive $V_G$. For $V_G < -20$ V, this enhanced field-assisted transport of the photogenerated minority carriers across the heterojunction is shown by the thicker blue arrows in Figure S12b. In contrast, for positive $V_G$, the net electric field that drives the photogenerated minority carriers is diminished.

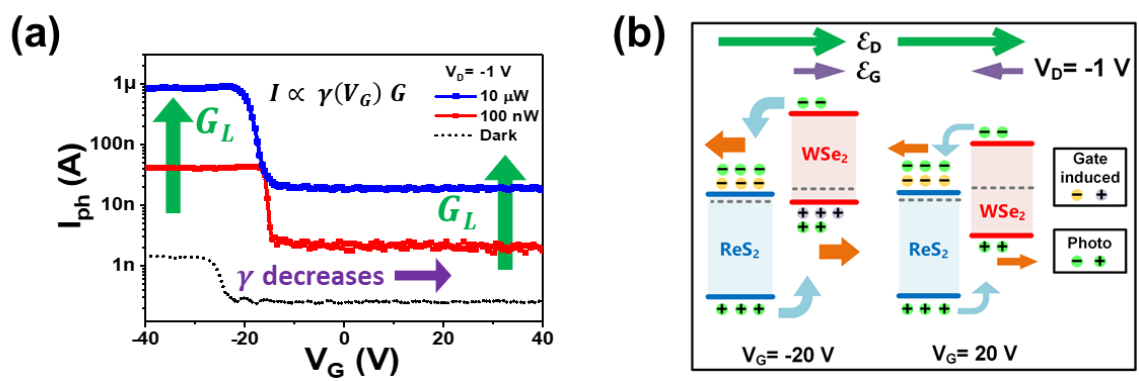

**Figure S13**: (a) Transfer characteristics of FL/FL $WSe_2/ReS_2$ heterostructure diode under reverse bias at 100 nW and 10 µW incident laser illumination. The dark characteristics are shown in dotted lines. (b) Representative band diagrams depicting the band alignments under reverse bias for $V_G = \pm 20$ V.

## S14: Transfer characteristics under illumination for forward bias ($V_D = 1$ V) (FL/FL structure)

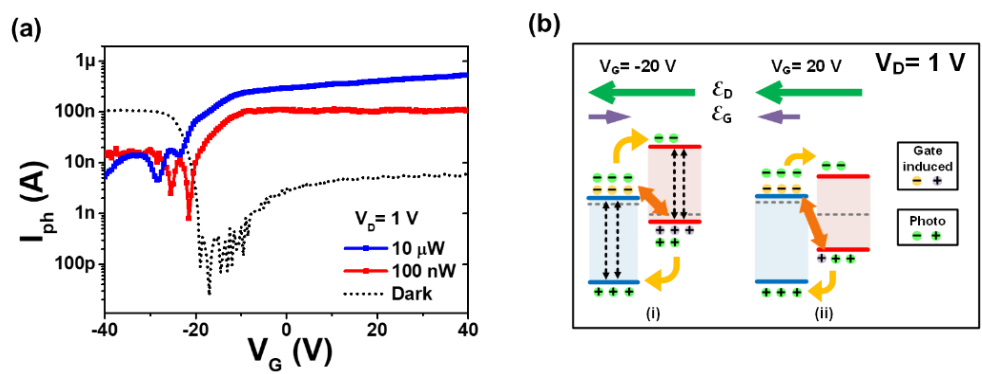

**Figure S14**: (a) Transfer characteristics of FL/FL $WSe_2/ReS_2$ heterostructure under forward bias at 100 nW and 10 µW incident laser illumination. The dark characteristics are shown in dotted lines. (b) Representative band diagrams depicting the band alignments under forward bias for $V_G = \pm 20$ V.

Dependence of the photocurrent on the gate voltage is shown in (a) and representative forward bias band diagrams at $V_G$ = -20 V and $V_G$ = 20 V are described in (b). A constant generation rate independent of the gate voltage is considered. The carriers induced due to the gate voltage as well as the photogenerated carriers are marked in (b). For $V_G < -20$ V, the increased number of carriers leads to enhanced recombination thereby lowering the photocurrent as compared to $V_G > -20$ V. Also, the electric field due to the source-drain bias is in the direction of the gate field for positive $V_G$.

## S15: Ideality factor of the WSe$_2$/ReS$_2$ photovoltaic device

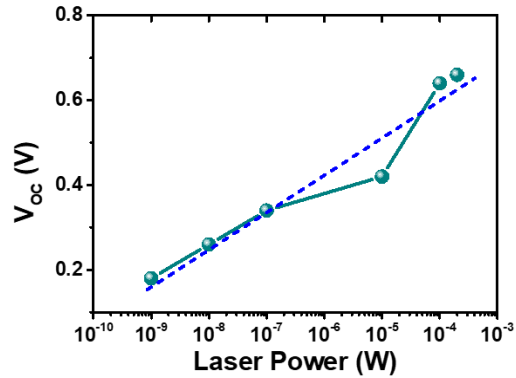

**Figure S15**: $V_{OC}$ vs. incident laser power for the FL/FL heterostructure. The ideality factor of the photovoltaic device is obtained from the slope of the curve.

From the Shockley diode equation for a photovoltaic cell, the ideality factor ($n_{id}$) can be determined under open circuit conditions by:

$$\frac{\partial V_{OC}}{\partial \ln P_{laser}} = n_{id} \frac{k_B T}{q} \qquad (1)$$

A $n_{id}= 1.5$ is obtained, implying both Langevin (direct recombination) and SRH (trap assisted recombination mechanisms) influence the current transport.

## S16: $V_{OC}$ dependence on gate voltage for FL/FL heterostructure

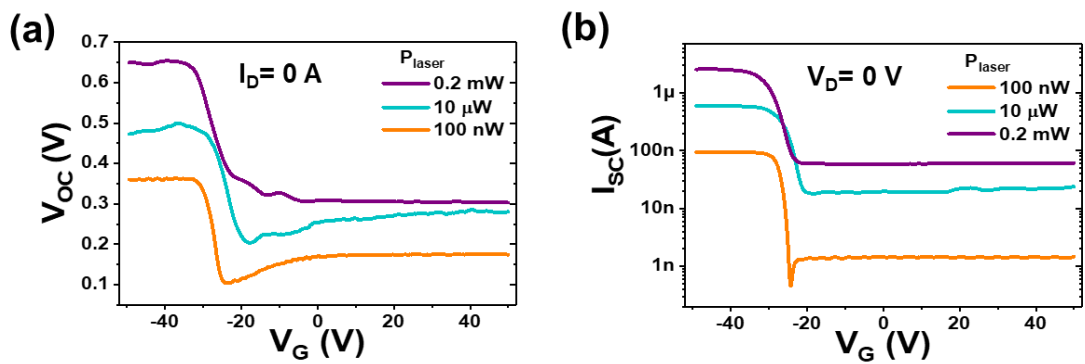

**Figure S16**: Gate voltage dependence of (a) $V_{OC}$ and (b) $I_{SC}$ at different illumination powers.

## S17: $V_{OC}$ dependence on flake thickness

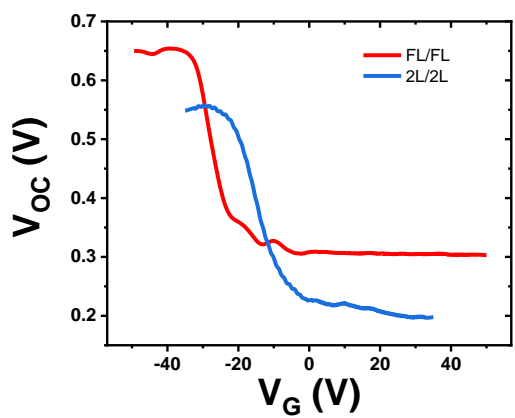

**Figure S17**: Gate voltage dependence of $V_{OC}$ for heterostructures of different thicknesses (FL/FL and 2L/2L).

The open circuit voltage depends on the dark current as well as the current under illumination. Since the photocurrent is higher for the FL/FL heterostructure, the $V_{OC}$ values are higher. However, the dependence of the $V_{OC}$ with back gate voltage is similar for both devices.